\documentclass[twocolumn]{aastex631}

\usepackage{natbib}
\usepackage{amsmath}
\usepackage{subfigure}
\setcitestyle{sort,comma}
\usepackage{appendix}
\usepackage{csvsimple}

\usepackage[version=4]{mhchem} 
\usepackage{seqsplit}

\providecommand{\apjl}{ApJL}
\providecommand{\apjs}{ApJS}
\providecommand{\aap}{A\&Aj}

\graphicspath{{./}{figures/}}

\begin{document}



\title{\large{Possible Evidence for the Presence of Volatiles on the Warm Super-Earth TOI-270\,b}}



\correspondingauthor{Louis-Philippe Coulombe}
\email{louis-philippe.coulombe@umontreal.ca}

\newcommand{\umontreal}{Department of Physics and Trottier Institute for Research on Exoplanets, Universit\'{e} de Montr\'{e}al, Montreal, QC, Canada}

\newcommand{\ucla}{Department of Earth, Planetary, and Space Sciences, University of California, Los Angeles, Los Angeles, CA, USA}

\author[0000-0002-2195-735X]{Louis-Philippe Coulombe} 
\affil{\umontreal}

\author[0000-0001-5578-1498]{Bj\"{o}rn Benneke} 
\affil{\ucla}
\affil{\umontreal}

\newcommand{\uwashington}{Department of Earth and Space Sciences/Astrobiology Program, University of Washington, Seattle, WA, USA}

\author[0000-0001-6878-4866]{Joshua Krissansen-Totton} 
\affil{\uwashington}

\author[0009-0005-6135-6769]{Alexandrine L'Heureux} 
\affil{\umontreal}

\newcommand{\uchicago}{Department of Astronomy \& Astrophysics, University of Chicago, Chicago, IL, USA}

\author[0000-0002-2875-917X]{Caroline Piaulet-Ghorayeb} 
\altaffiliation{E. Margaret Burbridge Postdoctoral Fellow}
\affil{\uchicago}

\author[0000-0002-3328-1203]{Michael Radica} 
\altaffiliation{NSERC Postdoctoral Fellow}
\affil{\uchicago}

\author[0000-0001-6809-3520]{Pierre-Alexis Roy} 
\affil{\umontreal}

\newcommand{\mpia}{Max-Planck-Institut f\"{u}r Astronomie, K\"{o}nigstuhl 17, D-69117 Heidelberg, Germany}

\author[0000-0003-0973-8426]{Eva-Maria Ahrer} 
\affil{\mpia}

\author[0000-0001-9291-5555]{Charles Cadieux}
\affil{\umontreal}

\newcommand{\sron}{SRON Netherlands Institute for Space Research, Niels Bohrweg 4, 2333 CA Leiden, The Netherlands}

\newcommand{\leiden}{Leiden Observatory, Leiden University, Einsteinweg 55, 2333 CC Leiden, The Netherlands}

\author[0000-0002-0747-8862]{Yamila Miguel}
\affil{\sron}
\affil{\leiden}

\author[0000-0002-0298-8089]{Hilke E. Schlichting}
\affil{\ucla}

\newcommand{\cab}{Centro de Astrobiología (CAB), CSIC-INTA, ESAC campus, Camino Bajo del Castillo s/n, 28692, Villanueva de la Cañada (Madrid), Spain}

\newcommand{\porto}{Instituto de Astrofísica e Ciências do Espaço, Universidade do Porto, CAUP, Rua das Estrelas, 4150-762 Porto, Portugal}

\author[0000-0003-4434-2195]{Elisa Delgado-Mena}
\affil{\cab}
\affil{\porto}

\author[0009-0005-9152-9480]{Christopher Monaghan} 
\affil{\umontreal}

\newcommand{\uclaAstro}{Department of Physics \& Astronomy, University of California Los Angeles, Los Angeles, CA 90095, USA}

\author[0009-0008-2624-0746]{Hanna Adamski}
\affil{\uclaAstro}

\newcommand{\uwisconsin}{Department of Astronomy, University of Wisconsin--Madison, Madison, WI 53706, USA}

\author[0009-0002-2380-6683]{Eshan Raul}
\affil{\uwisconsin}

\newcommand{\mcmaster}{Department of Physics \& Astronomy, McMaster University, Hamilton, ON, Canada}

\author[0000-0001-5383-9393]{Ryan Cloutier}
\affil{\mcmaster}

\newcommand{\oxford}{Department of Physics, University of Oxford, Parks Rd, Oxford, OX1 3PU, UK}

\author[0000-0002-9258-5311]{Thaddeus D. Komacek}
\affil{\oxford}

\author[0000-0003-4844-9838]{Jake Taylor}
\affil{\oxford}

\newcommand{\heidelberg}{Department of Physics and Astronomy, Heidelberg University, Im Neuenheimer Feld 226, D-69120 Heidelberg, Germany}

\author[0009-0007-9356-8576]{Cyril Gapp}
\affil{\mpia}
\affil{\heidelberg}

\author[0000-0002-1199-9759]{Romain Allart}
\altaffiliation{SNSF Postdoctoral Fellow}
\affil{\umontreal}

\newcommand{\riogrande}{Departamento de Física Teórica e Experimental, Universidade Federal do Rio Grande do Norte, Campus Universitário, Natal, RN, 59072-970, Brazil}

\newcommand{\geneve}{Observatoire de Genève, Département d’Astronomie, Université de Genève, Chemin Pegasi 51, 1290 Versoix, Switzerland}

\author[0000-0002-7613-393X]{François Bouchy}
\affil{\geneve}

\author[0000-0001-5578-7400]{Bruno L. Canto Martins}
\affil{\riogrande}

\author[0000-0003-4166-4121]{Neil J. Cook}
\affil{\umontreal}

\author[0000-0001-5485-4675]{René Doyon} 
\affil{\umontreal}

\newcommand{\newcastle}{School of Information and Physical Sciences, University of Newcastle, Callaghan, NSW, Australia}

\author[0000-0001-5442-1300]{Thomas M. Evans-Soma}
\affil{\newcastle}
\affil{\mpia}

\newcommand{\grenoble}{Univ. Grenoble Alpes, CNRS, IPAG, 38000 Grenoble, France}

\author[0009-0005-1139-3502]{Pierre Larue}
\affil{\grenoble}

\newcommand{\iac}{Instituto de Astrof\'{\i}sica de Canarias, c/ V\'ia L\'actea s/n, 38205
La Laguna, Tenerife, Spain\label{iac}
Departamento de Astrof\'{\i}sica, Universidad de La Laguna, 38206 La Laguna, Tenerife, Spain}

\author[0000-0002-3814-5323]{Alejandro Su{\'a}rez Mascare{\~n}o}
\affil{\iac}

\author[0000-0003-3191-2486]{Joost P. Wardenier}
\affil{\umontreal}


\begin{abstract}

The search for atmospheres on rocky exoplanets is a crucial step in understanding the processes driving atmosphere formation, retention, and loss.
Past studies have revealed the existence of planets interior to the radius valley with densities lower than would be expected for pure-rock compositions, indicative of the presence of large volatile inventories which could facilitate atmosphere retention. 
Here we present an analysis of the JWST NIRSpec/G395H transmission spectrum of the warm ($T_\mathrm{eq,A_B=0} = 569$\,K) super-Earth TOI-270\,b ($R_\mathrm{p}$ = 1.306\,$R_\oplus$), captured alongside the transit of TOI-270\,d. The JWST white light-curve transit depth updates TOI-270\,b's density to $\rho_\mathrm{p}$ = $3.7\pm0.5$\,g/cm$^3$, inconsistent at 4.4$\sigma$ with an Earth-like composition. Instead, the planet is best explained by a non-zero, percent-level water mass fraction, possibly residing on the surface or stored within the interior.
The JWST transmission spectrum shows possible spectroscopic evidence for the presence of this water as part of an atmosphere on TOI-270\,b, favoring a H$_2$O-rich steam atmosphere model over a flat spectrum ($\ln\mathcal{B}=0.3-3.2$, inconclusive to moderate), with the exact significance depending on whether an offset parameter between the NIRSpec detectors is included. We leverage the transit of the twice-larger TOI-270\,d crossing the stellar disk almost simultaneously to rule out the alternative hypothesis that the transit-light-source effect could have caused the water feature in TOI-270\,b's observed transmission spectrum. Planetary evolution modeling furthermore shows that TOI-270\,b could sustain a significant atmosphere on Gyr timescales, despite its high stellar irradiation, if it formed with a large initial volatile inventory.

\end{abstract}



\section{Introduction}\label{sec:intro}

The Solar System presents a clear dichotomy between the gas giants, hosting primordial H$_2$/He-dominated atmospheres, and the terrestrial planets, hosting either high mean-molecular-weight (MMW) secondary atmospheres or no atmosphere at all. One of the main legacies of the \textit{Kepler} survey is the discovery of a similar phenomenon, in which the occurrence of planets show a bimodal distribution composed of super-Earths and sub-Neptunes, two populations peaking at $R_\mathrm{p}\sim$ 1.3\,$R_\oplus$ and $\sim$ 2.4\,$R_\oplus$, respectively, with a gap in occurrence rate reaching a minimum at 1.8\,$R_\oplus$ \citep{Fulton2017,Van_eylen2018,Fulton2018}.
This gap is generally thought to be the result of photoevaporative and core-powered mass-loss processes \citep{Owens2013,Lopez2013,Sheng2014,Chen2016,Ginzburg2018,Gupta2019}, with super-Earths expected not to retain any significant H$_2$/He-dominated atmospheres. Another proposed explanation is that the radius valley is the outcome of planet formation processes alone. \citet{Venturini2020} showed that the bimodality of the planetary radii distribution is reproduced through pebble accretion of more rocky or icy material depending on formation location. Additionally, the significance and position of the radius gap has been observed to depend on the host star spectral type, with M-dwarf stars exhibiting a higher occurrence of planets within the radius gap compared to early-type stars \citep{Cloutier2020,Ho2024,Parc_2024,Venturini_2024}. Planets within the radius gap have been found to show densities consistent with the presence of significant amounts of volatiles.
An analysis of a sample of 34 small planets orbiting M dwarfs identified a distinct population of objects: “water worlds”, planets with significant ($>$10\%) amounts of volatiles (e.g., H$_2$O) by mass \citet{Luque_2022}. However, a recent analysis of a twice-larger sample of planets observed no statistical evidence for the presence of such a population \citep{dainese2025robuststatisticalevidencepopulation}. Ultimately, the limited size and precision of the current sample make these results sensitive to statistical analysis methods and interior structure assumptions \citep{Parviainen2024}.

The TOI-270 planetary system \citep{Gunther_2019} is a prime example of the transition between super-Earths and sub-Neptunes. It contains three planets spanning both sides of the radius valley \citep[$R_\mathrm{p,b}$ $\sim$ 1.3, $R_\mathrm{p,c}$ $\sim$ 2.3, and $R_\mathrm{p,d}$ $\sim$ 2.0\,$R_\oplus$;][]{Kaye2022} orbiting a quiet (by M-dwarf standards) M3V-type host star, proving a valuable test subject to study the outcome of planet formation, evolution, and migration.


The innermost planet of the system, TOI-270\,b \citep[$R_p = 1.28_{-0.04}^{+0.05} \,R_\oplus$,][]{Kaye2022}, sits directly in the radius regime of super-Earths. Past studies found its measured density to be consistent with an Earth-like composition \citep{Van_Eylen_2021,Kaye2022}, and it is expected to have rapidly shed off any primordial H$_2$/He it might have accreted throughout its formation \citep{Owen_2017}. It is possible, however, that it accreted a significant amount of volatiles during its formation or through late-stage impacts, and could thus host a secondary atmosphere sustained by outgassing \citep{Kite2020}. Late-type stars, with their small radii and low effective temperatures, present a prime opportunity to characterize the atmosphere of small, rocky exoplanets through transmission and thermal emission observations. Owing to its relatively high equilibrium temperature ($T_\mathrm{eq,A_B=0}$ = 581\,K) and the radius of its host star \citep[$R_*$ = 0.378\,$R_\odot$;][]{Van_Eylen_2021}, TOI-270\,b has a predicted feature amplitude of $\sim$40\,ppm for a water-dominated atmosphere (eq. 2 of \citealt{Kreidberg_2018}, assuming that five scale heights are probed), which is in principle within the reach of the JWST.

While the size of TOI-270 proves advantageous for atmospheric studies, the possible presence of unocculted spots and faculae on its surface, which are expected on active stars such as M dwarfs, may lead to the introduction of stellar contamination in transmission spectroscopy via the Transit Light-Source (TLS) effect \citep{Rackham2018,Rackham_2019}. Thus far, this source of contamination has hindered efforts to characterize rocky planets via transmission spectroscopy  \citep[e.g.,][]{Moran_2023,Lim2023,May2023,radica2024promiseperilstellarcontamination,Rathcke2025}. This is especially problematic for M-dwarf stars, as water absorption features in the stellar spectrum can create the illusion of planetary water absorption features. Notably, \citet{Moran_2023} and \citet{May2023} found evidence for water features in the JWST/NIRSpec G395H transmission spectra of GJ 486\,b and GJ 1132\,b, respectively. These features could be explained either by a planetary atmosphere signal or contamination from unocculted star spots, with the two scenarios being strongly degenerate at the wavelengths covered by the observations (3--5\,$\mu$m). JWST/MIRI thermal emission observations later revealed both of these planets to show dayside flux measurements consistent with bare rocks \citep{xue2024jwstthermalemissionterrestrial,mansfield2024thickatmosphereterrestrialexoplanet}, suggesting that the observed features in transmission were possibly due to stellar contamination after all. The thermal emission measurements could not rule out thin atmospheres with moderate amounts of water, however, leaving some margin for the presence of tenuous atmospheres on these planets.

An analysis of the NIRSpec/G395H spectrum of TOI-270\,b was presented in \citet{holmberg2024possiblehyceanconditionssubneptune}, where they found that both stellar contamination and H$_2$-rich atmosphere models are preferred at more than 3-$\sigma$ significance over a featureless spectrum.
They further found that the H$_2$-rich atmosphere model is tentatively preferred (2.7$\sigma$) over the stellar contamination model when imposing Gaussian priors on the TLS parameters measured from TOI-270\,d. Few details are given as to the possible composition of the H$_2$-rich atmosphere scenario, however, with only a potential preference for absorption from OCS being mentioned. Furthermore, no retrievals were performed fitting simultaneously for TLS and an atmosphere model, which can significantly impact the atmospheric inferences \citep{piauletghorayeb2025strictlimitspotentialsecondary}.

In this work, we present an updated density for the super-Earth planet TOI-270\,b using the more precise radius constraints from its JWST NIRSpec/G395H transit,
revealing the planet to be significantly underdense compared to an Earth-like composition. We also perform an in-depth atmospheric analysis of the transmission spectrum, presenting possible evidence of a water absorption feature in its transmission spectrum even when accounting for the TLS effect, which we constrain at high precision using the transmission spectrum of TOI-270\,d. In section \ref{sec:obs_red}, we present the JWST NIRSpec/G395H observations, along with the data reduction and light-curve fitting methodology. We also explain the derivation of stellar abundances from high-resolution near-infrared spectroscopy with the NIRPS spectrograph (Near-InfraRed Planet Searcher). We describe the bulk composition analysis in section \ref{sec:interior}, atmospheric analysis and results in sections \ref{sec:retrieval} and \ref{sec:resu}, respectively, and the atmospheric evolution analysis in section \ref{sec:evolution}. Finally, we discuss the implications of our results in section \ref{sec:discussion} and summarize our work in section \ref{sec:conclusion}.

\section{Observations and Data Analysis}\label{sec:obs_red}

As part of a large JWST survey of sub-Neptunes and water worlds (GO 4098; P.I.: Benneke \& Evans-Soma), we observed on Oct 13, 2023, a transit of the sub-Neptune TOI-270\,d with the NIRSpec/G395H instrument \citep{Jakobsen_2022,Birkmann_2022}. Independent analyses of this visit are presented in \citet{benneke2024jwstrevealsch4co2} and \citet{holmberg2024possiblehyceanconditionssubneptune}. These observations also captured a transit of the innermost planet of the system, super-Earth TOI-270\,b, the analysis of which is the focus of this work. We detail our data reduction, light-curve preparation, and light-curve fitting methodology in the subsections below.

\subsection{exoTEDRF Data Reduction}
We perform a reduction of the TOI-270\,b/d time series observations using \texttt{exoTEDRF} \citep{Feinstein_2023, radica_2023_awesome, radica_2024_exotedrf} to complement the \texttt{Eureka!} \citep{Bell2022} and \texttt{Tiberius} \citep{Kirk2017,Kirk2021}  analyses presented in \citet{benneke2024jwstrevealsch4co2}. We closely follow the procedure presented in \citet{schmidt2025comprehensivereanalysisk218bs} and \citet{Ahrer_2025}, and outline the major steps here. 

The Stage 1 corrections include subtraction of the superbias and dark frames, linearity correction, and cosmic ray detection, for which we use a time-domain rejection algorithm \citep{radica_2024_muted} to identify and flag any pixels which deviate by more than 10$\sigma$ from a median stack of the seven surrounding frames. We also correct the column-correlated 1/$f$ noise and background signal at the group level following best practices for NIRSpec observations \citep[e.g.,][]{alderson_2023_early, alderson_2024_jwst}. For this step, we mask all pixels within nine pixels of the trace center, as well as any clear outliers, and subtract the median of each detector column to simultaneously remove the background and 1/$f$ contributions. Stage 1 ends with ramp fitting. 

In Stage 2, we repeat the background and 1/$f$ subtraction to remove any residual column-correlated noise. However, we note that adding this step has a negligible impact on the final extracted spectra, indicating that our initial group-level correction was sufficient. We then flag and interpolate any remaining bad pixels using a 15$\sigma$ spatial (i.e., outliers relative to surrounding pixels in a given frame) and 10$\sigma$ temporal (i.e., outliers relative to surrounding frames in time) outlier threshold. We then perform a principal component analysis on the 2D detector frames themselves in order to identify detector trends which may impact the extracted light curves \citep{Coulombe2023, radica_2024_muted,Coulombe2025highlyreflectivewhiteclouds}. However, we find that the detector was extremely stable throughout the course of the TSO, with only a minor sub-pixel drift in the vertical position of the trace over time. Finally, we locate the positions of the spectral traces on each detector using the \texttt{edgetrigger} algorithm \citep{radica_2022_applesoss} and extract the stellar spectra using a simple box aperture with a full width of eight pixels.  

\subsection{Light-Curve Preparation}

We perform additional corrections to the light curves extracted with \texttt{exoTEDRF} before proceeding with light-curve fitting. 
Temporal outliers are corrected for in each spectral bin by clipping any count that is more than 3$\sigma$ away from the running median.
We also correct for “bad” spectral bins by computing the point-to-point scatter in time of all bins and flagging any bin whose measured scatter is 3$\sigma$ away from the running median in wavelength. The count values of all flagged bins are then replaced by the value of the running median in wavelength space. We produce two white light curves, one for each NRS detector, and also produce the spectroscopic light curves using a fixed resolving power of $R=200$, resulting in 53 and 60 bins for NRS1 and NRS2, respectively, for a total of 113 spectral bins. We repeat the same procedure for the \texttt{Eureka!} and \texttt{Tiberius} reductions presented in \citet{benneke2024jwstrevealsch4co2} to produce three independent spectra.

\begin{table*}[hbt!]
\caption{Derived orbital parameters from the joint fit of our TOI-270\,b JWST NIRSpec/G395H NRS1 and NRS2 white light curves extracted using \texttt{exoTEDRF}. Stated values correspond to the median and $\pm1\sigma$ uncertainties. We use the stellar radius from \citet{Kaye2022} ($R_*$ = $0.380\pm0.008$\,$R_\odot$) and the effective temperature from \citet{Van_Eylen_2021} ($T_{\mathrm{eff}}$ = $3506\pm70$\,K) to derive the planetary radius and equilibrium temperature. The planetary density is derived using the mass from \citet{Kaye2022} ($M_\mathrm{p}=1.48\pm0.18$\,$M_\oplus$).}
\centering
\begin{tabular}{lccc}
\hline
\hline
Parameter      &       \citet{Van_Eylen_2021}       & NIRSpec/G395H         \\
\hline
Mid-transit time $T_0$ [BJD - 2400000.5]     &  58387.09505 $\pm$ 0.00074  &  60221.740084 $\pm$ 0.000035   \\
Planet-to-star radius ratio $R_p/R_\mathrm{s}$   &  0.02920 $\pm$ 0.00069  &  0.03144 $\pm$  0.00010 \\
Semi-major axis $a/R_\mathrm{s}$   &  18.18$_{-0.52}^{+0.57}$  & 19.09$_{-0.69}^{+0.33}$ \\
Impact parameter $b$      &  0.19 $\pm$ 0.12  &  0.20 $\pm$ 0.13 \\
\hline
\hline
Derived &  & \\
\hline
Planetary radius $R_\mathrm{p}$ [R$_\oplus$]  & 1.206 $\pm$ 0.039 & 1.306 $\pm$ 0.028 \\
Inclination $i$ [deg] & 89.39 $\pm$ 0.37 & 89.41$_{-0.43}^{+0.40}$ \\
Semi-major axis $a$ [au] &  0.03197$\pm$0.00022 & 0.03362$_{-0.00128}^{+0.00099}$ \\
Equilibrium temperature $T_\mathrm{eq,A_B = 0}$ [K] & 581 $\pm$ 14 & 569$_{-13}^{+14}$ \\
Equilibrium temperature $T_\mathrm{eq,A_B = 0.3}$ [K] & 532 $\pm$ 13 & 521$_{-12}^{+13}$ \\
Mean Planetary Density $\rho_\mathrm{p}$ [g/cm$^3$] & 4.97 $\pm$ 0.94 & $3.7\pm0.5$ \\ 
Mean Stellar Density $\rho_\mathrm{s}$ [g/cm$^3$] &  7.20 $\pm$ 0.63 & 11.30$_{-0.32}^{+0.29}$ \\ 
\hline
\label{table:wlc_fit}
\end{tabular}
\end{table*}

\subsection{Light-Curve Fitting}

We separate the light-curve fitting process into two distinct steps following a methodology similar to that of \citet{benneke2024jwstrevealsch4co2}. First, we perform a joint fit of our NRS1 and NRS2 white light curves to constrain the orbital and physical parameters of TOI-270\,b and d. Second, we proceed with the spectroscopic light-curve fitting, where we fix the orbital parameters to the best-fit values from the white light-curves fit. We describe these two steps in detail below.

\subsubsection{White Light-Curve Fit}\label{sec:wlc_fit}

To constrain the orbital parameters of TOI-270\,b, we jointly fit the two white light curves produced by summing all wavelengths of the NRS1 (2.86--3.72\,$\mu$m) and NRS2 (3.82--5.17\,$\mu$m) detectors from the \texttt{exoTEDRF} reduction. We compute the running median of both light curves (using a window of 5 integrations) and clip any data point that is more than 3.5$\sigma$ away from the median. We keep free the orbital parameters ($T_\mathrm{0}$, $R_\mathrm{p}/R_\mathrm{s}$, $a/R_\mathrm{s}$, $b$) of TOI-270\,b and TOI-270\,d, assuming large uniform priors. We fix the period of planets b and d to $P_\mathrm{b}$ = 3.3601538 and $P_\mathrm{d}$ = 11.379573\,days \citep{Van_Eylen_2021}, respectively, and assume circular orbits ($e$ = 0). We use the quadratic limb-darkening law and fit the limb-darkening coefficients (LDCs) [$u_1,u_2$] considering large uniform priors ($\mathcal{U}[-3,3]$), as recommended in \cite{Coulombe2024LD}, sharing the coefficients between the two planets. As in \citet{benneke2024jwstrevealsch4co2}, we sum the transit light-curve models of both planets, simulated with \texttt{batman} \citep{Kreidberg_2015}, to produce the astrophysical model. We consider a linear trend for the systematics model and fit for the photometric scatter. The parameter space is explored using the Markov chain Monte Carlo ensemble sampler \texttt{emcee} \citep{Foreman_Mackey_2013}. We iterate for 50,000 steps, using 80 walkers (4 per free parameter) and discard the first 30,000 steps as burn-in. The systematics-corrected white light curves (in which we removed the contribution from TOI-270\,d), along with their best fits, are shown in Figure \ref{fig:lightcurves}. The constraints on the orbital parameters of TOI-270\,b are presented in table \ref{table:wlc_fit}.

The measured semi-major axis and impact parameter for TOI-270\,b are consistent with the values from \citet{Van_Eylen_2021} within their 1$\sigma$ errors. We also measure planet-to-star radius ratios from the NRS1 ($R_\mathrm{p}/R_\mathrm{s}$ = 0.03164 $\pm$ 0.00013) and NRS2 ($R_\mathrm{p}/R_\mathrm{s}$ = 0.03117 $\pm$ 0.00015) white light curves, which we combine into a single measurement of the planetary radius using the weighted mean and uncertainty of the two values, also accounting for the correlation between the two parameters, resulting in a planet-to-star radius ratio of $R_\mathrm{p}/R_\mathrm{s}$ = 0.03144 $\pm$ 0.00010. Assuming the stellar radius from \citet{Kaye2022} of $R_* = 0.380\pm0.008\,R_\odot$, our measured $R_\mathrm{p}/R_\mathrm{s}$ translates to a planetary radius of $R_\mathrm{p}=1.306\pm0.028\,R_{\oplus}$ (Figure \ref{fig:Mp_vs_Rp}), consistent at less than 1$\sigma$ with the values presented in \citealt{Kaye2022} ($R_\mathrm{p}=1.28^{+0.05}_{-0.04}\,R_{\oplus}$) and \citealt{holmberg2024possiblehyceanconditionssubneptune} ($R_\mathrm{p}=1.302\pm0.028\,R_{\oplus}$), and at 2$\sigma$ with the value of \citealt{Van_Eylen_2021} ($R_\mathrm{p}=1.206\pm0.039\,R_{\oplus}$).

Our constraints on the planet's semi-major axis together with the orbital period reported by \citet{Van_Eylen_2021} correspond to a measurement of the stellar density ($\rho_\mathrm{s}$), via Kepler's third law \citep{Seager2003,winn2014transitsoccultations}

\begin{equation}
    \rho_\mathrm{s} \approx \frac{3\pi}{GP^2} \left(\frac{a}{R_\mathrm{s}}\right)^3.
\end{equation}

\medskip 
\noindent 
To constrain this value, we repeat the white light-curve analysis described above, this time directly fitting for the stellar density ($\mathcal{U}[1,20]$\,g/cm$^3$, shared between planets b and d), rather than for the semi-major axes. Our fit yields a value of $\rho_\mathrm{s} = 11.30_{-0.32}^{+0.29}$\,g/cm$^3$, in agreement with and more precise than the value of $10.63\pm0.74$\,g/cm$^3$ derived in \citet{Kaye2022}. While our measured density is inconsistent with that of \citet{Van_Eylen_2021} ($7.20\pm0.63$\,g/cm$^3$), this is likely due to an erratum as their measured stellar radius and mass correspond to a density of $10.11\pm0.93$\,g/cm$^3$. We do not re-derive the stellar mass and radius from our density measurement, however, as potentially non-zero eccentricities for planets b and d, known to lead to overestimation of the stellar density \citep{Kipping2010}, would need to be accounted for.

\subsubsection{Spectroscopic Light-Curve Fits}

\begin{figure}
\begin{center}
\vspace{-2.5mm}
\includegraphics[width=0.45\textwidth]{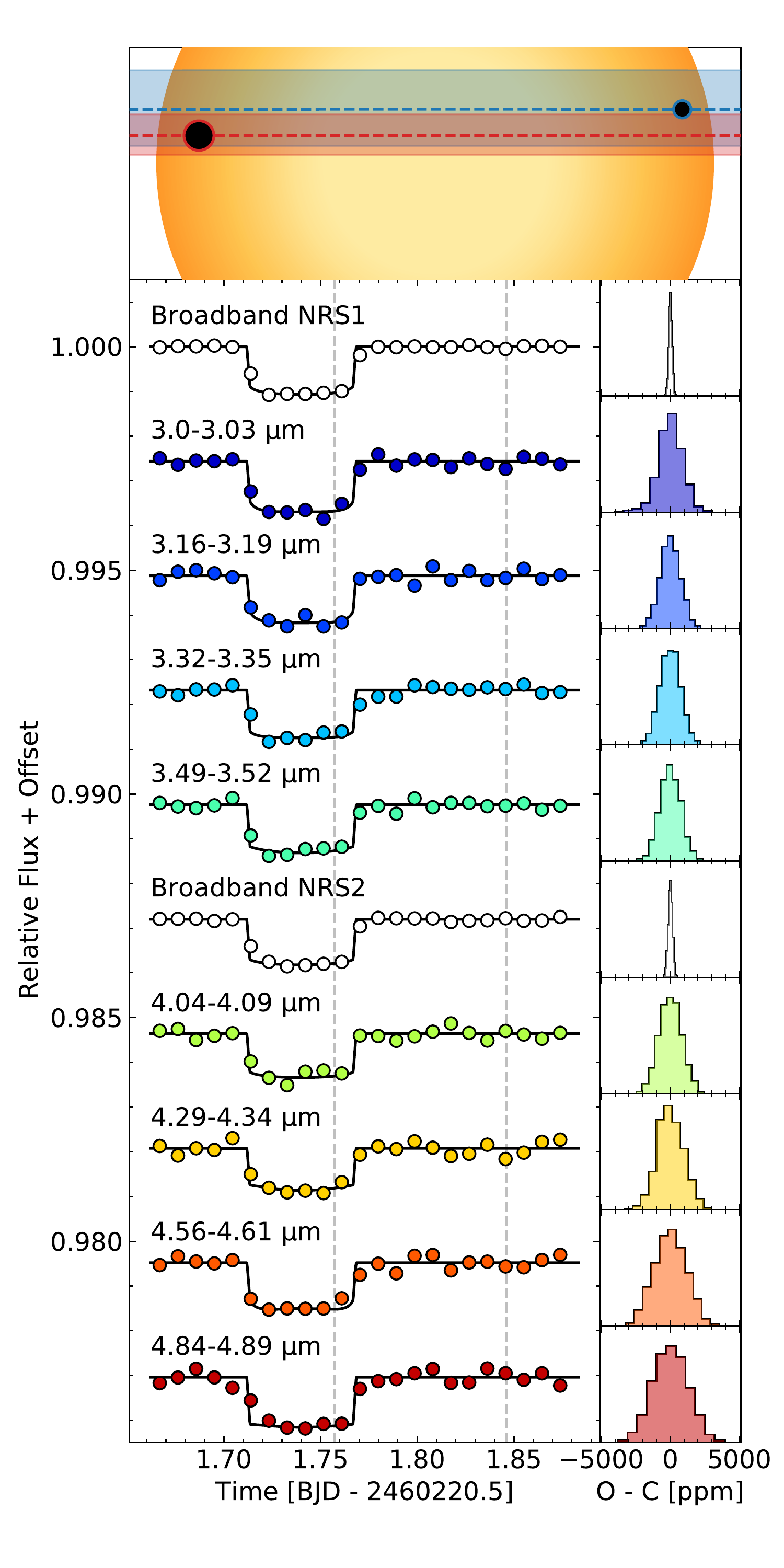}
\end{center}
\vspace{-11mm}\caption{Broadband and spectroscopic light-curve fits of the NIRSpec/G395H transit of TOI-270\,b, extracted with \texttt{exoTEDRF}. The top panel is a schematic of TOI-270\,b (blue) and d (red) as they transit across the stellar disk at time $t = 2460222.265$\,BJD, illustrating their trajectories (dashed lines) and the 1$\sigma$ confidence region corresponding to their orbital inclination uncertainties. The broadband light-curve fits (white) are shown for the NRS1 and NRS2 detectors, along with eight systematics-corrected spectroscopic light curves (colored dots) and their best-fit transit models (black lines). The light curves are plotted with a relative flux offset and are binned in time in groups of 75 data points (corresponding to 13.6 minute bins). The best-fit model transit of TOI-270\,d, whose position is indicated by the gray dashed lines, partially overlaps with that of TOI-270\,b and is removed from the data and best-fit model for visual clarity. All light curves are well-behaved and show Gaussian distributions in their residuals (computed from the non-binned light curves, right).} 
\label{fig:lightcurves}
\end{figure}

We proceed with the spectroscopic light-curve fitting by fixing the orbital parameters of TOI-270\,b and TOI-270\,d to their best-fit values from the white light-curve fit ($T_\mathrm{0,b}$ = 2460222.240112\,BJD, $a_\mathrm{b}/R_\mathrm{s}$ =  19.33, $b_\mathrm{b}$ = 0.12, $T_\mathrm{0,d}$ = 2460222.301636\,BJD, $a_\mathrm{d}/R_\mathrm{s}$ = 42.92, $b_\mathrm{d}$ = 0.094). We then follow the same procedure as for the white light curves, fitting each spectral bin individually. We keep free the planet-to-star radius ratio of both planets, assume that they share the same limb-darkening coefficients per wavelength, and consider a linear trend for the systematics model. We sample the parameter space for 10,000 steps per spectral bin, using four walkers per free parameter, and discard the first 6,000 steps as burn-in. Examples of systematics-corrected spectroscopic light curves from \texttt{exoTEDRF} are shown in Figure \ref{fig:lightcurves} and the resulting transmission spectrum of TOI-270\,b is shown in Figures \ref{fig:spec_vs_mods} and \ref{fig:reducs_specs}. Spectra obtained from the \texttt{Eureka!} and \texttt{Tiberius} reductions using the same fitting methodology are shown in Figure \ref{fig:reducs_specs}.

\subsection{Correlated Noise and Impact of Data Reduction Methodology}

We find that the fitted NRS1 and NRS2 white light-curve scatters are significantly higher for the \texttt{Eureka!} reduction compared to \texttt{Tiberius} and \texttt{exoTEDRF} (Fig. \ref{fig:RMS_plots}). However, we observe a “noise plateau” beyond bin sizes of $\sim$30--40\,minutes and $\sim$10\,minutes for the NRS1 and NRS2 detectors, respectively, beyond which the three reductions exhibit similar residual RMS values. The existence of this plateau could be caused by the presence of astrophysical noise or uncorrected systematics common to all reductions. The root-mean-square (RMS) of the residuals as a function of bin size for the spectroscopic light curves also deviate from the expected behavior of Poisson noise, although less notably than the white light curves given their lower signal-to-noise.

Comparing the transmission spectra from the three reductions, we find good agreement between the \texttt{Tiberius} and \texttt{exoTEDRF} spectra, with the \texttt{Eureka!} spectrum diverging from the two other reductions at the reddest wavelengths ($>$4.5\,$\mu$m) of the NRS2 detector. To ensure that there is no correlation between neighboring spectral channels, we compute the covariance matrices of the residuals from the spectroscopic light-curve fits for all three reductions (Figure \ref{fig:spec_corr}). In the scenario that there is no cross-talk between spectral bins, we expect all elements of the covariance matrix to be null except for its diagonal. 
Upon visual inspection of the covariance matrices, we find significant correlation between neighboring spectral channels for the \texttt{Eureka!} reduction. Quantitatively, the average maximum-normalized covariance between neighboring bins is of $0.13\pm0.03$, compared to $0.02\pm0.03$ and $0.01\pm0.03$ for \texttt{Tiberius} and \texttt{exoTEDRF}, respectively. This is possibly indicative of leftover 1/$f$ noise in the light curves, which can introduce correlation between neighboring columns and would also explain the larger scatter observed for that reduction. We thus do not consider the spectrum extracted with \texttt{Eureka!} for the remainder of this work. Given the consistency between the \texttt{Tiberius} and \texttt{exoTEDRF} reductions, as well as the slightly lower transit depth errors of the latter, we proceed with the \texttt{exoTEDRF} spectrum for our atmospheric analysis. We also repeat a portion of our atmospheric analysis on the \texttt{Tiberius} spectrum to assess the robustness of our results to data reduction assumptions.



\subsection{NIRPS TOI-270 Stellar Spectrum Analysis}\label{sec:nirps_analysis}

As part of the NIRPS Guaranteed Time Observation \citep[NIRPS-GTO;][Bouchy et al., submitted]{ArtigauSPIE2024} sub-program focused on the characterization of exoplanet atmospheres (Prog ID 112.25P3.001; PI: Bouchy \& Doyon), four transits of TOI-270\,c and two of TOI-270\,d were observed between November 19, 2023, and November 4, 2024.
NIRPS \citep{WildiSPIE2022, ArtigauSPIE2024} is a near-infrared (0.98--1.80\,$\mu$m) fiber-fed high-resolution spectrograph installed in parallel with HARPS \citep{MayorHARPS2003, PepeHARPS2004} on the ESO 3.6-meter telescope at La Silla Observatory, Chile.
Observations of TOI-270 were conducted in High Efficiency mode (HE; $R\approx70,000$)
, each comprising 22 to 33 consecutive exposures of 400\,s, with median signal-to-noise ratio (SNR) per pixel of $\sim60$ in the center of the $H$ band. The observations are reduced with \texttt{APERO} \citep{cookAPEROPipelinEReduce2022} v0.7.292.
Here, we use the 173 individual telluric-corrected spectra to construct a single high-resolution spectrum of TOI-270 from which the stellar abundances are determined.

To measure the abundances of specific chemical species, we perform a series of fits on individual spectral lines following the methodology of \citet{jahandarComprehensiveHighresolutionChemical2024, jahandarChemicalFingerprintsDwarfs2025}. The lines are fitted through $\chi^2$ minimization to state-of-the-art high resolution synthetic models of M dwarf stars from the PHOENIX ACES library \citep{Husser_2013}. The models are convolved to the resolution of NIRPS and have a fixed $T_\mathrm{eff}$ of 3500\,K, following the measurement of \citet{Van_Eylen_2021}. We do not use the effective temperature derived from NIRPS, as we find significant discrepancy between the effective temperatures measured in the $Y$ and $J$ bands compared to the $H$ band ($T_{\mathrm{eff}, YJ}=3564\pm30$\,K and $T_{\mathrm{eff,} H}=3731\pm20$\,K). This discrepancy is possibly due to systematic errors in the stellar models, as well as the lack of $K$-band coverage, which was found to be essential in breaking the metallicity-temperature degeneracy in SPIRou data \citep{jahandarComprehensiveHighresolutionChemical2024}. Finally, we note that the stellar abundance uncertainties are likely underestimated as we are not marginalizing over the uncertainty on $T_\mathrm{eff}$. Table \ref{table:stellarparams} records the abundances of all the chemical species detected in this TOI-270 dataset, including key refractory elements such as Fe, Mg and Si, which compose the bulk of planetary cores and mantles. We obtain a solar iron-to-magnesium ratio \citep[$\mathrm{Fe/Mg}=0.81\pm0.22$, the solar value being $0.79\pm0.10$,][]{Asplund2009}.
The overall metallicity and its uncertainty are computed by taking the average abundances of all the measured elements and their standard deviation divided by $\sqrt{N-1}$, $N$ being the number of measurements. This approach, yielding $\mathrm{[M/H]}=-0.13\pm0.07$, allows to better capture the dispersion of the element abundances, reducing the weight put on oxygen due to the high number of OH lines \citep{jahandarComprehensiveHighresolutionChemical2024}.

\section{Planetary Bulk Composition Analysis}\label{sec:interior}

Our planet-to-star radius ratio of TOI-270\,b obtained from its JWST NIRSpec transit results in an updated planetary radius measurement that is 1.4 and 1.6 times more precise than the values of \citet{Van_Eylen_2021} and \citet{Kaye2022}, respectively. Motivated by this increase in precision, we perform an interior structure analysis for the planet.

\subsection{Interior Structure Analysis}

We use the \texttt{smint} python package \citep{Piaulet_2021,Piaulet_2022} to model the measured planetary radius and mass of TOI-270\,b, assuming that the structure of the planet consists of three distinct layers: an iron-rich core, a silicate-rich mantle, and a hydrosphere, following the prescription described in \citet{Aguichine_2021}. The hydrosphere is composed of a steam atmosphere atop a supercritical water layer. \texttt{smint} takes as input the planetary mass ($M_\mathrm{p}$), irradiation temperature ($T_\mathrm{irr}$), water mass fraction (WMF, $f_\mathrm{H_2O}$), and core mass fraction of the interior (CMFI, $f'_\mathrm{core}$), and interpolates over the \citet{Aguichine_2021} model grid to compute the corresponding planetary radius. The CMFI is computed considering only the core and mantle of the planet, in contrast with the core mass fraction (CMF, $f_\mathrm{core}$) which also accounts for the mass of the hydrosphere. The interpolator is then coupled to the affine-invariant ensemble sampler \texttt{emcee} \citep{Foreman_Mackey_2013} to marginalize over the $f'_\mathrm{core}$--$T_\mathrm{irr}$--$M_\mathrm{p}$ parameter space and infer the range of WMF values that are consistent with the observed planet radius.

When there is no a priori knowledge of the CMFI, it is common to marginalize over $f'_\mathrm{core}$ by assuming a uniform prior $\mathcal{U}[0,1]$. However, because an earth-like composition corresponds to a CMFI of $f'_{\mathrm{core},\oplus}$ = 0.325 \citep{Aguichine_2021}, a uniform prior on $f'_\mathrm{core}$ allots twice as much prior volume to iron-rich scenarios ($f'_{\mathrm{core}}>f'_{\mathrm{core},\oplus}$) compared to iron-poor ($f'_{\mathrm{core}}<f'_{\mathrm{core},\oplus}$). To prevent this, we instead marginalize directly over the planetary iron-to-magnesium ratio [Fe/Mg]/[Fe/Mg]$_\odot$, allowing for values ranging from 0.1--10 times the solar ratio ($\mathcal{U}$[-1,1]).
Assuming that the core is made of pure iron and the mantle of pure enstatite (MgSiO$_3$), the CMFI can be described as

\begin{equation}\label{eq:CMFI}
f'_\mathrm{core} = \left[1+\left(\frac{\mathrm{Fe}}{\mathrm{Mg}}\right)\frac{\mu_\mathrm{MgSiO_3}}{\mu_\mathrm{Fe}}\right]^{-1},
\end{equation}

\medskip 
\noindent 
where $\mu_\mathrm{Fe}/\mu_\mathrm{MgSiO_3}$ is the iron-to-enstatite MMW ratio. Using this reparameterized form, we fit for our measured planetary radius considering a uniform prior for $f_\mathrm{H_2O}$ ($\mathcal{U}[0,1]$), and Gaussian priors for $T_\mathrm{irr}$ ($\mathcal{N}[569,14^2]$\,K, Table \ref{table:wlc_fit}) and $M_\mathrm{p}$ ($\mathcal{N}[1.48,0.18^2]$\,$M_\oplus$, \citealt{Kaye2022}).

If, rather than being agnostic of the core mass fraction of the planet, we presume that it is directly related to the measured Fe/Mg ratio of its host star, we can further constrain the water mass fraction of TOI-
270\,b. This assumption is based on Solar System observations which find that refractory elemental ratios of planetary materials match that of the Sun \citep{Lodders2003}. The same trend is expected for exoplanets \citep{Dorn2015,Hinkle2018}, and is supported by population studies \citep{brinkman2024compositionsrockyplanetsclosein}. We note, however, that processes such as collisions \citep[][hypothesized to be the cause of Mercury's overdensity]{Benz1988} can impact the overall core-to-mantle mass ratio, and thus the stellar abundances are not a perfect one-to-one proxy for planetary abundance ratios. 
Our measurement of the stellar iron-to-magnesium molar ratio of $\mathrm{Fe/Mg}=0.81\pm0.22$ \citep[comparable to the solar value of $0.79\pm0.11$,][]{Asplund2009} thus corresponds to a core mass fraction of the interior of $f'_\mathrm{core}=0.407\pm0.074$ (equation \ref{eq:CMFI}), which we then impose as a Gaussian prior in \texttt{smint} when fitting for the water mass fraction. Additionally, we perform a fit assuming a Gaussian prior based on the solar value ($\mathcal{N}$[0.412,0.032$^2$]).

\subsection{Density and Internal Structure of TOI-270\,b}

\begin{figure}[h]
\begin{center}
\includegraphics[width=0.48\textwidth]{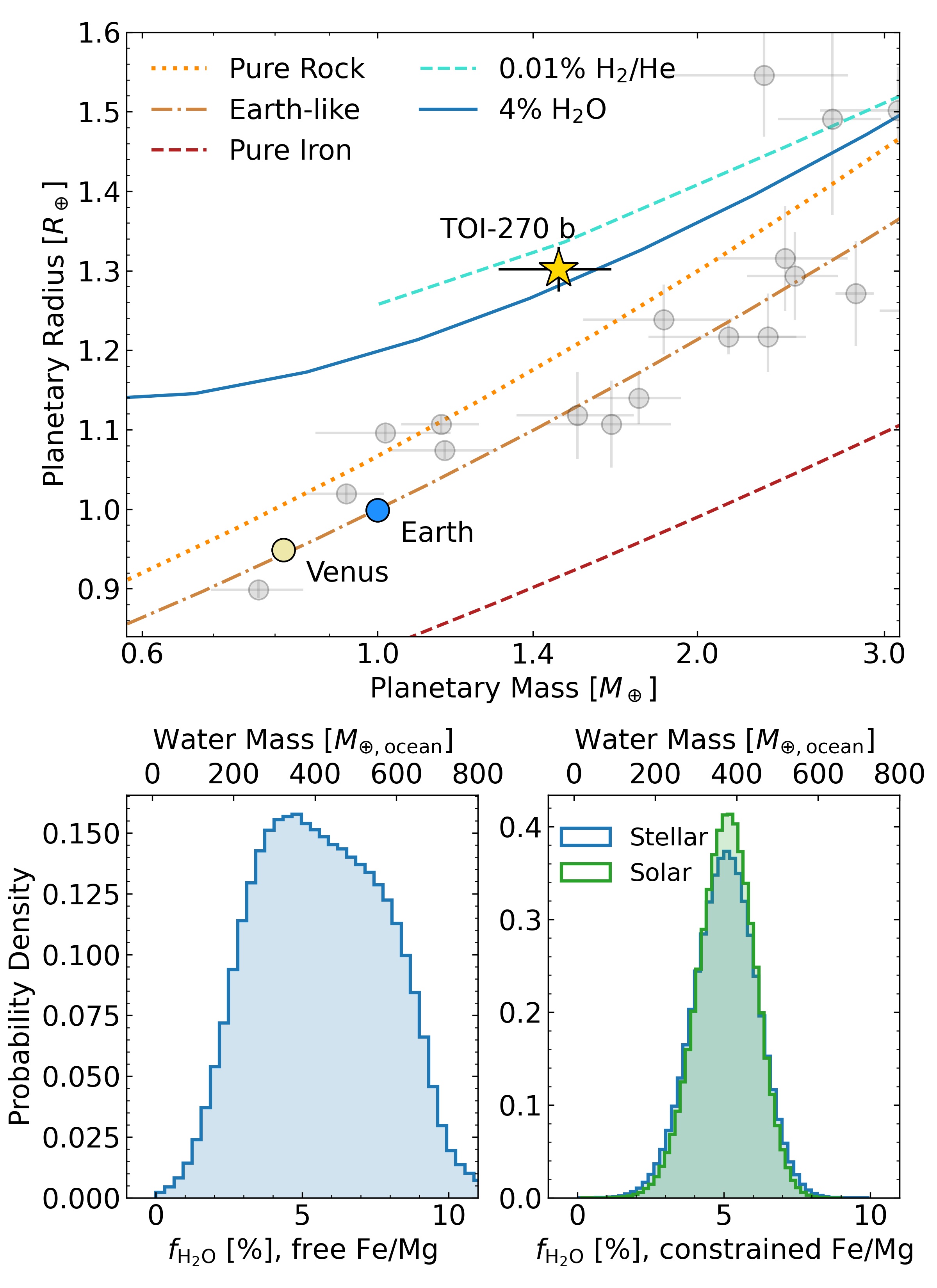}
\end{center}
\vspace{-5mm}\caption{\textbf{Top:} Mass radius diagram for all planets with equilibrium temperatures (A$_\mathrm{B}$ = 0, $f$ = 1) below 1,000\,K, radii below 1.6\,R$_\oplus$, and masses measured at more than 20\% precision. The gold star indicates the position of TOI-270\,b, using our measured radius and the mass of \citet{Kaye2022}. Isocomposition model curves from \citet{Zeng2016} are shown for pure iron (red), Earth-like (33\% iron; brown), and pure rock (orange) scenarios. In contrast with planets that are in a similar mass range, TOI-270\,b is inconsistent with an Earth-like composition at 4.4$\sigma$. While consistent with a pure-rock composition at 2.4$\sigma$, its radius is best explained by 3--6\% water by mass when considering the irradiated ocean model from \citet{Aguichine_2021} (blue line). The planet is also slightly below the 0.01\% H$_2$/He curve (light blue line), the lowest value modeled in the atmosphere models of \citet{Lopez_2014}. \textbf{Bottom Left:} Inferred water mass fraction ($f_\mathrm{H_2O}$) probability posterior distribution when assuming a log-uniform prior on the planetary iron-to-magnesium ratio to sample the CMFI. \textbf{Bottom Right:} Inferred water mass fractions when assuming the stellar (blue) and solar (green) Fe/Mg values as priors for the CMFI.} 
\label{fig:Mp_vs_Rp}
\end{figure}

Combined with the mass of $1.48\pm0.18$\,M$_\oplus$ from \cite{Kaye2022}, our measured radius of $1.302\pm0.028$\,R$_\oplus$ corresponds to a mean density of $3.7\pm0.5$\,g/cm$^3$ for TOI-270\,b (Table \ref{table:wlc_fit}), in agreement within their 1$\sigma$ uncertainties with the measurements of \citet{Van_Eylen_2021} ($4.97\pm0.94$\,g/cm$^3$) and \citet{Kaye2022} ($3.89\pm0.66$\,g/cm$^3$). When compared to the isocomposition mass-radius curves from \citet{Zeng2016}, we find that TOI-270\,b is inconsistent with an Earth-like composition (33\% iron) at 4.4$\sigma$ and is also 2.4$\sigma$ above the pure rock (100\% mantle) curve (Figure \ref{fig:Mp_vs_Rp}). We note that the difference in interpretation regarding the interior structure of TOI-270\,b with that of \citet{Kaye2022}, who found its density to be consistent with a pure rock, is not caused by a significant difference in the measured density but rather by a difference in the interior structure models used. Indeed, the bulk composition curves considered in \citet{Gunther_2019} and \citet{Kaye2022} were taken from \citet{Fortney2007}, which, in contrast with \citet{Zeng2016}, does not account for density variations throughout the core and mantle. 

Additionally, the \texttt{smint} structure modeling shows that the density of TOI-270\,b is best-explained by the presence of \%-level water by mass on TOI-270\,b, with an inferred WMF of 5.6$_{-2.2}^{+2.5}$\% when assuming a log-uniform prior on the planetary Fe/Mg ratio (Figure \ref{fig:Mp_vs_Rp}). 
The range of possible WMF values is reduced when assuming the stellar and solar Fe/Mg values as priors for the CMFI of the planet, resulting in constraints of $5.1\pm1.1$\% and $5.1\pm1.0$\%, respectively.
However, as discussed in section \ref{sec:nirps_analysis} the uncertainty on the stellar Fe/Mg value is possibly underestimated, which could in turn lead to optimistic errorbars on the WMF inferred from the stellar prior. Furthermore, we note that the region of the parameter space for WMF values between 0 and 10\% is sampled in \texttt{smint} by interpolating between the models of \citet{Zeng2016} and \citet{Aguichine_2021}. Without finer sampling of the \citet{Aguichine_2021} models in the 0--10\% WMF range, it is difficult to assess the accuracy of this interpolation and, consequently, the exact WMF of TOI-270\,b. Nevertheless, our results show the WMF to be below 10\% and more than 4$\sigma$ away from 0 when assuming the solar and stellar priors.

One possibility is that, rather than TOI-270\,b being underdense, the assumed stellar radius and mass are incorrect. Although these two quantities cannot be inferred directly from transit light curves, our constraint on the stellar density of TOI-270 ($\rho_s$ = $11.30_{-0.32}^{+0.29}$\,g/cm$^3$) is consistent with the mass and radius measurements of \citet{Van_Eylen_2021} and \citet{Kaye2022}, derived using apparent magnitude, parallax, and spectral constraints.
We thus conclude that the stellar parameters used to derive the planetary radius and mass of TOI-270\,b are indeed valid.

\section{Atmospheric Analysis}\label{sec:retrieval}


We model our transmission spectrum of TOI-270\,b with varying levels of chemical assumptions, from self-consistent atmosphere models to free chemistry retrievals, and various treatments of the TLS effect, as described in detail below. All analyses are performed on the $R = 200$ \texttt{exoTEDRF} spectrum, and a subset of the retrievals is repeated on the \texttt{Tiberius} spectrum for comparison.

\subsection{Self-Consistent Atmosphere Models}

We compare our measured transmission spectrum to atmosphere models produced using the SCARLET framework \citep{Benneke2012,Benneke2013,benneke2015strictupperlimitscarbontooxygen,Benneke_2019_K218b,Benneke_2019}. Four self-consistent models are produced assuming equilibrium chemistry, for atmospheric metallicities of 1, 100, 300, and 1,000 times solar.
The carbon-to-oxygen ratio is fixed to the solar value \citep[0.55, ][]{Asplund2009}, and we assume full heat recirculation ($f$ = 1) and zero Bond albedo. The temperature-pressure profiles for these models are computed in a self-consistent manner, iteratively updating the temperature and chemistry of sixty evenly-spaced atmospheric layers ($P=10^{4}$--$10^{-8}$\,bar) until the atmosphere reaches radiative and chemical equilibrium (following the Rybicki method outlined in \citealt{Hubeny2014}). The absorption of CH$_4$ \citep{Yurchenko2017}, H$_2$O \citep{Polyansky_2018}, CO \citep{Hargreaves2019}, CO$_2$ \citep{Yurchenko2020}, NH$_3$ \citep{Coles2019}, HCN \citep{Harris2006}, H$_2$S \citep{Azzam2016}, PH$_3$ \citep{Sousa_Silva_2014}, Na, and K \citep{Piskunov1995,Burrows2003}, as well as the collision-induced absorptions from H$_2$--H$_2$ and H$_2$--He \citep{borysow2002}, are considered to simulate the spectra. Moreover, we produce transmission spectra for 100\,bar isothermal ($T(p) = 569$\,K; Table \ref{table:wlc_fit}) atmospheres composed of 100\% H$_2$O, CH$_4$, CO$_2$, and NH$_3$, molecules that have significant absorption features over the NIRSpec/G395H wavelength range. Our suite of forward models is shown in Figure \ref{fig:spec_vs_mods}.

\subsection{Atmosphere-Only Retrieval}\label{sec:atm_only_ret}

We perform atmospheric retrievals on the transmission spectrum of TOI-270\,b using SCARLET, which takes as input the temperature-pressure and molecular abundance profiles, as well as a set of parameters for a given cloud prescription, and returns the wavelength-dependent transit depth of a planet.
The atmospheric model is then coupled to a Bayesian inference algorithm to solve the inverse problem and constrain the properties of the atmosphere.

We model the atmosphere as a set of 60 layers with equal spacing in log-space from 10$^{4}$--10$^{-10}$\,bar.  
The temperature-pressure profile of the planet is considered to be constant with altitude and is fit assuming a Gaussian prior ($\mathcal{N}$[569,100$^2$]\,K) centered at the equilibrium temperature of the planet (for zero Bond albedo and full heat redistribution), with a wide standard deviation of 100\,K.
For the atmosphere constituents, we model a mixture of water and a fill gas.
The abundance of water is fit considering the centered-log-ratio transform of \citet{Benneke2012}, with a prior of $\mathcal{U}[L/n,L(1-n)/n]$ ($L = -10$ is the lower limit in log space of the abundances, $n = 2$ is the number of species) for the transformed volume mixing ratio $\xi_i$.
The water cross-sections are computed from the POKAZATEL line list \citep{Polyansky_2018} using HELIOS-K \citep{Grimm2021}, and we consider the collision-induced absorption of H$_2$-H$_2$ following the coefficients of \citet{borysow2002}. We also fit for an effective surface pressure  $\log P_\mathrm{eff,surf}$ ($\mathcal{U}$[-8,2]\,$\log$[bar]), which acts to reproduce the presence of either a gray cloud top or solid surface at a given pressure level.

For the fill gas, we consider two distinct scenarios: First, a scenario where molecular hydrogen (H$_2$) makes up the rest of the atmosphere composition and, second, a scenario where the fill gas is a fictitious element X, whose MMW $\mu_\mathrm{X}$ is a free parameter. The fill gas X has no opacity, and its only impact on the spectrum is through the effect of its MMW on the atmospheric scale height. We assume a uniform prior for the MMW of species X ($\mathcal{U}[2,44]$\,AMU), ranging from that of H$_2$ to CO$_2$, such that we remain agnostic of the possible MMW of inert/unconstrained species making up the rest of the atmosphere composition of TOI-270\,b.

Finally, since offsets between the two NRS detectors have been observed in previous NIRSpec/G395H datasets \citep[e.g.,][]{alderson_2024_jwst,Wallack2024}, we run all retrievals described in this work twice: once assuming no offset, and once including an offset as a free parameter ($\delta_\mathrm{NRS,1,2}$, $\mathcal{U}$[-250,250]\,ppm). Unless stated otherwise, the results shown in the rest of this analysis are obtained from the retrievals assuming no offset, and the impact of its inclusion on the atmospheric inferences is discussed in section \ref{sec:offset}.
The parameter space is explored using the nested sampling python package \texttt{nestle} \citep{Skilling2004,Skilling2006,Barbary2014} with 30,000 live points, a tolerance factor of 0.5 on the log-evidence, and the single ellipsoid sampling method.

\subsection{Free TLS effect Retrieval}\label{sec:atm+free_tls}

We also perform retrievals accounting for the transit light source effect, following the methodology described in \citet{Fournier_Tondreau_2023} and \citet{Piaulet_Ghorayeb_2024}. We consider only the effect of spots, which are able to mimic water absorption features, and fit for a spot covering fraction ($f_\mathrm{spot}$, $\mathcal{U}$[0,1]), spot temperature contrast ($\Delta T_\mathrm{spot}$, $\mathcal{U}$[$-$1000,0]\,K), and stellar photosphere temperature ($T_\mathrm{*,phot}$, $\mathcal{N}$[3506,70$^2$]\,K; \citeauthor{Van_Eylen_2021}). We do not consider faculae in the retrievals as no evidence for their presence in the spectrum of TOI-270\,d was found \citep{benneke2024jwstrevealsch4co2}.
PHOENIX stellar models \citep{Husser_2013} are used to produce the contamination spectra, interpolated at a surface gravity $\log g$ = 4.872 and metallicity [Fe/H] = $-0.2$ \citep{Van_Eylen_2021}. The models are computed for a range of effective temperatures $T_\mathrm{eff}\in$[2506,4506]\,K in 20\,K increments using the \texttt{MSG} module \citep{Townsend2023}. The stellar contamination is fit simultaneously with the atmosphere model described in subsection \ref{sec:atm_only_ret}. Because we want to fit only for the TLS in this case -- assuming that the planetary spectrum is flat -- we fit for the abundance of He, which is spectrally inactive, instead of H$_2$O, and consider species X for the fill gas. 
We again explore the parameter space using 30,000 live points.

\subsection{Joint Atmosphere \& constrained TLS effect Retrieval}\label{sec:atm_TLScon_ret}

\begin{figure}
\includegraphics[width=0.48\textwidth]{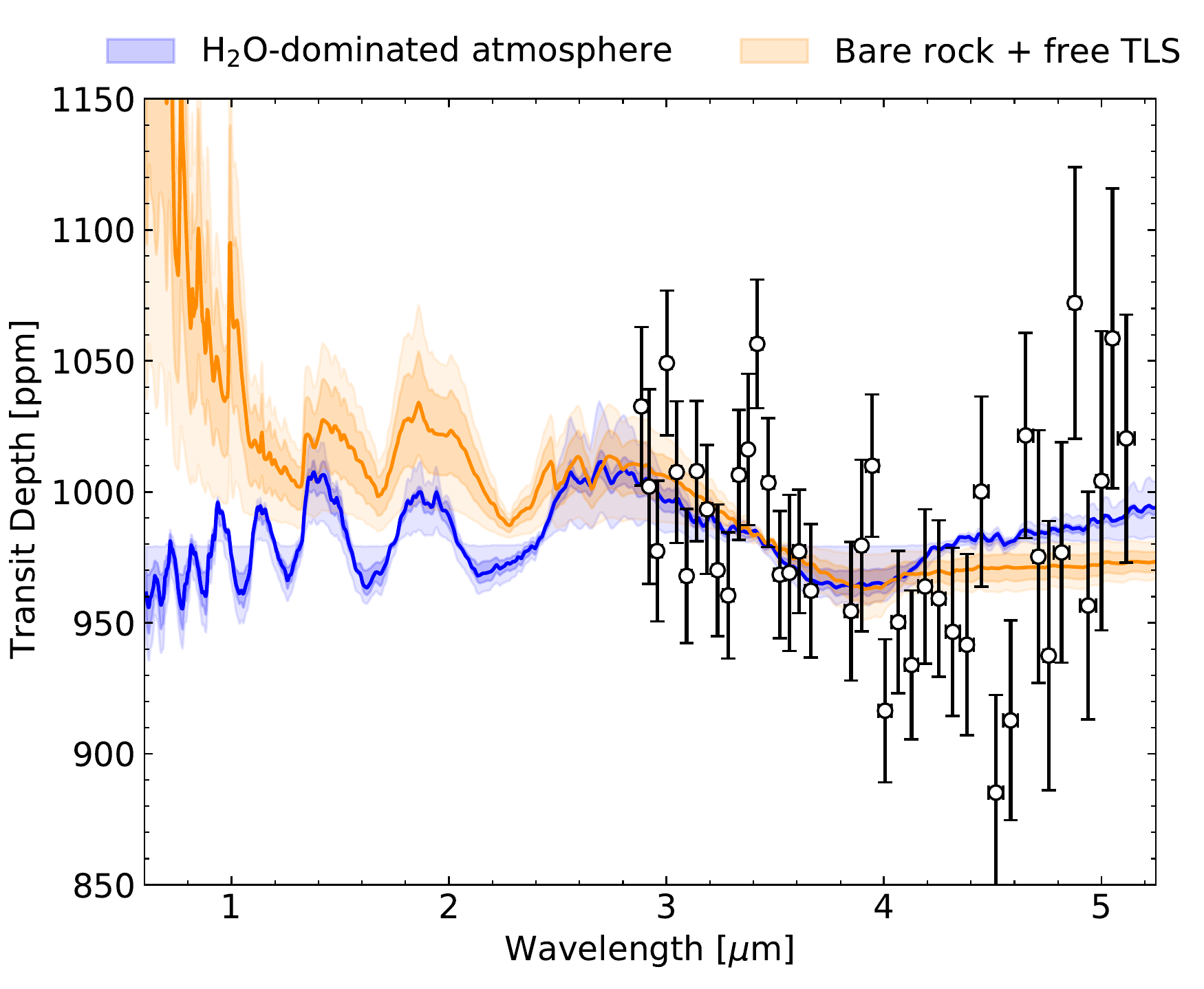}
\caption{The NIRSpec/G395H transmission spectrum of TOI-270\,b (black circles, binned at a fixed resolving power of R = 80 for visual clarity) is well fit by either an H$_2$O-rich atmosphere or stellar contamination. The median fit of the atmosphere-only (X + H$_2$O) and bare-rock + TLS retrievals, along with their 1/2-$\sigma$ confidence regions, are shown in blue and orange, respectively. The atmosphere and TLS effect models are similar over the NIRSpec/G395H wavelength range, with the difference between these two effects being more easily discernable towards optical wavelengths.}
\label{fig:atm_vs_tls}
\end{figure}

As shown in Figure \ref{fig:atm_vs_tls}, there is a degeneracy between the presence of a water absorption feature from a possible atmosphere and from unocculted stellar heterogeneities, mainly due to the lack of short-wavelength coverage where these two effects are typically best distinguished \citep[e.g.,][]{Lim2023,Fournier_Tondreau_2023,cadieux2024transmissionspectroscopyhabitablezone,radica2024promiseperilstellarcontamination}. Previous observations of super-Earths with NIRSpec/G395H have faced a similar problem \citep[e.g.,][]{Moran_2023,May2023}. However, one further option is available to us: we can leverage the transit of TOI-270\,d, observed in the same visit \citep{benneke2024jwstrevealsch4co2}, to further constrain the TLS effect. For a given planet, its wavelength-dependent effective transit depth ($\delta_\mathrm{p,eff}$) can be described as

\begin{equation}
\delta_\mathrm{p,eff}(\lambda) = \delta_\mathrm{atm}(\lambda)\epsilon_\mathrm{het}(\lambda),
\end{equation}

\medskip
\noindent
where $\delta_\mathrm{atm}$ is the transit depth of the planet and its atmosphere, and $\epsilon_\mathrm{het}$ is the contamination factor introduced by unocculted stellar spots and faculae \citep{MacDonald2022}. Thus, the absolute deviation ($\Delta\delta_\mathrm{p,eff}$) from the planetary atmosphere spectrum that is introduced by the TLS effect is equal to

\begin{equation}
\Delta\delta_\mathrm{p,eff}(\lambda) = \delta_\mathrm{atm}(\lambda) (\epsilon_\mathrm{het}(\lambda)-1) \approx \left(\frac{R_\mathrm{p}}{R_\mathrm{s}}\right)^2 (\epsilon_\mathrm{het}(\lambda)-1).
\end{equation}

\medskip
\noindent
This signifies that, the larger the planet-to-star ratio of a planet is, the more it is sensitive to stellar contamination. Applying this logic to the Solar System, one could possibly use a transit of Jupiter to constrain the contamination from the Sun's heterogeneities at a hundred times the precision than could be done with Earth, on the condition that they cross similar transit chords. As TOI-270\,d is 1.7 times larger than TOI-270\,b \citep{benneke2024jwstrevealsch4co2}, a stellar contamination feature of 40\,ppm in the spectrum of planet b would correspond to a feature of 115\,ppm for planet d.

\begin{figure}[h]
\begin{center}
\includegraphics[width=0.48\textwidth]{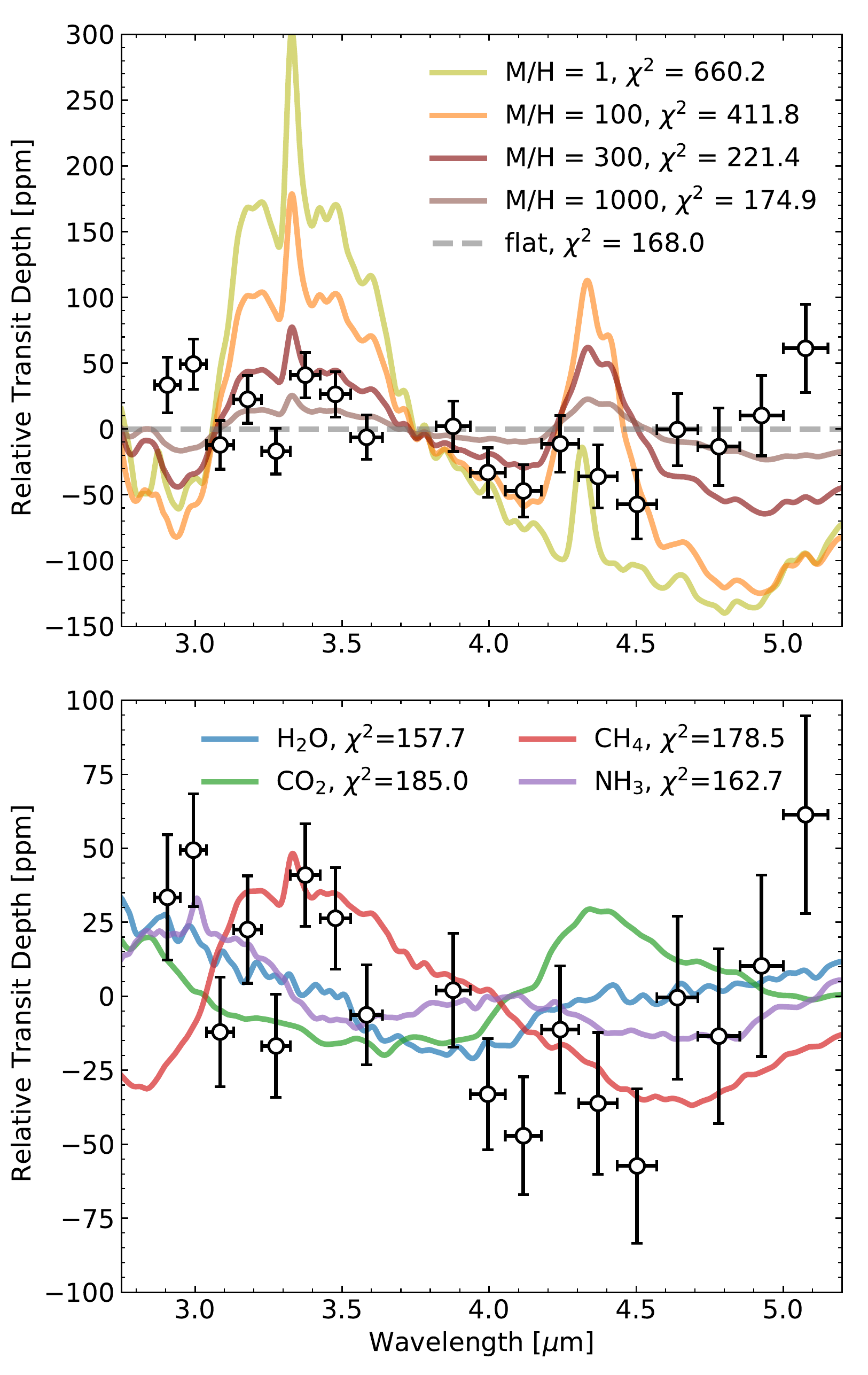}
\end{center}
\vspace{-5mm}\caption{\textbf{Top:} Comparison of our measured transmission spectrum, binned at a fixed resolving power of R = 35 for visual clarity, to cloudless self-consistent models produced using SCARLET and assuming a solar carbon-to-oxygen ratio \citep[C/O = 0.55,][]{Asplund2009} and atmosphere metallicities of 1 (yellow), 100 (orange), 300 (maroon), and 1,000 times solar (brown).
The relatively small amplitude of the spectrum of TOI-270\,b is inconsistent with any clear H$_2$/He-rich atmosphere with atmosphere metallicities below 300 times solar. Chi-square $\chi^2$ values between the models and the full-resolution ($R=200$, $N=113$ data points)
spectrum are shown for comparison. \textbf{Bottom:} Comparison of TOI-270\,b with 100\,bar pure H$_2$O (blue), CH$_4$ (red), CO$_2$ (green), and NH$_3$ (purple) atmospheres. The rise in transit depth at the shorter wavelengths of NIRSpec/G395H is best reproduced by the absorption features of water.} 
\label{fig:spec_vs_mods}
\end{figure}

To leverage this effect, we run free-chemistry retrievals on the NIRSpec/G395H transmission spectrum of TOI-270\,d, following the same methodology described in \citet{benneke2024jwstrevealsch4co2}, while also fitting for the TLS effect. Only spots are considered, and retrievals are run with and without accounting for an offset between NRS1 and NRS2. While we use the \texttt{Tiberius} spectrum for this portion of the analysis, as it is the reduction considered for the atmospheric analysis in \citet{benneke2024jwstrevealsch4co2}, we note that retrievals performed on the \texttt{exoTEDRF} of TOI-270\,d spectrum produce virtually the same constraints on the TLS parameters.  

We then repeat the retrievals described in Section \ref{sec:atm_only_ret}, this time using the constraints on the stellar contamination parameters from the retrieval of TOI-270\,d as a prior for our joint atmosphere and TLS effect retrieval of TOI-270\,b. The prior on the TLS effect is applied by taking the samples of $f_\mathrm{spot}$ and $\Delta T_\mathrm{spot}$ from TOI-270\,d and performing a 2D kernel density estimation (KDE). In the retrieval, we then interpolate over the natural logarithm of the KDE for a given set of TLS effect parameters and multiply the corresponding KDE value to the prior. This is similar to the method employed in \citet{holmberg2024possiblehyceanconditionssubneptune}, where they imposed truncated Gaussian priors on the spot covering fraction and spot contrast for their retrieval on the spectrum of TOI-270\,b. The benefit of the KDE method, however, is that the strong correlation between $f_\mathrm{spot}$ and $\Delta T_\mathrm{spot}$ (Fig. \ref{fig:atm_vs_tls}) is conserved. Because we cannot implement our KDE prior in the prior transform of \texttt{nestle}, we instead run these retrievals using \texttt{emcee}. We run a total of 100 chains 
for 20,000 steps, producing our posteriors from the last 40\% of the total steps and discarding the rest as burn-in. 

When using the constraints on the TLS parameters from TOI-270\,d for the atmospheric retrievals on TOI-270\,b, we make the assumption that both planets share the same transit chord. This assumption is justified by the fact that they show consistent impact parameters \citep[$b_\mathrm{b}$ = $0.20\pm0.13$, Table \ref{table:wlc_fit} and $b_\mathrm{d}$ = 0.101$^{+0.070}_{-0.069}$,][]{benneke2024jwstrevealsch4co2}, as illustrated in Figure \ref{fig:lightcurves}. Moreover, there is no evidence of spot/faculae crossing events in the transit light curves of both planets, indicating they both crossed “clean” transit chords and should thus share similar out-of-chord stellar surfaces.

\section{Atmospheric Retrieval Results}\label{sec:resu}

\subsection{Comparison with Self-Consistent Models}\label{sec:SCmods}
Comparison of our transmission spectrum with self-consistent clear, solar-C/O, H$_2$/He-rich atmospheres indicates that, for atmospheric metallicities below 1,000 times solar, the amplitude of the expected molecular features are much larger than the observed scatter in the transit depth measurements (Fig. \ref{fig:spec_vs_mods}). We compute the preference/rejection significance (through the Bayes factor $\mathcal{B}$ -- the odds ratio between two models) for these models relative to a flat line via the difference in their Bayesian Information Criterion ($\Delta\mathrm{BIC}$). The BIC depends on the maximum log likelihood of the model ($\mathcal{L}_\mathrm{max}$), its number of free parameters ($k$), and the number of data points ($N$ = 113): $\mathrm{BIC}$ = $-2\ln\mathcal{L}_\mathrm{max} + k \ln N$ \citep{Schwarz1978,Liddle_2007}.
The Bayes factor $\mathcal{B}$ can then be approximated as $\ln\mathcal{B} \approx -\Delta\mathrm{BIC}/2$ following \citet{Kass1995}.
The significance at which a model is preferred/rejected over another is then assessed using Jeffreys' scale (Appendix B of \citet{Jeffreys1939}, also presented in \citealt{Trotta_2008} and \citealt{Benneke2013}), where $\ln\mathcal{B}_{01}<1.0$ is inconclusive, $1.0\leq\ln\mathcal{B}_{01}<2.5$ is weak, $2.5\leq\ln\mathcal{B}_{01}<5.0$ is moderate, and $\ln\mathcal{B}_{01}\geq5.0$ is strong.
For all models, we compute the minimum chi-square (listed in Fig. \ref{fig:spec_vs_mods}) that is achieved when allowing for a free DC offset of the model (one free parameter, $k=1$) using the minimize function of the SciPy Python package. The maximum likelihoods $\mathcal{L}_\mathrm{max}$ are then computed from these chi-square values.

When comparing a straight line fit to the self-consistent models, we measure Bayes factors of $\ln\mathcal{B}=$246.1, 121.9, 26.7, and 3.5 for atmospheric metallicities of 1, 100, 300, and 1,000 times solar, respectively. Below atmospheric metallicities of 300 times solar, the resulting amplitude of the spectrum is too large to match the observed spectrum, resulting in these models being strongly ruled out.
For a solar-composition atmosphere under equilibrium chemistry conditions, methane and carbon dioxide are expected to be the main absorbers at the temperatures of TOI-270\,b.
These two species, however, produce a worse fit to the data (higher $\chi^2$) than a straight line for 100\,bar pure CO$_2$ ($\ln\mathcal{B}=-8.5$) and CH$_4$ ($\ln\mathcal{B}=-5.3$) atmospheres. Contrastingly, we find that a 100\,bar pure H$_2$O model is preferred ($\ln\mathcal{B}=5.2$) when compared to a straight line model. When varying the surface pressure of the pure H$_2$O model to 1 and 10\,bar, we find that the chi-square is virtually unchanged, as the amplitude of the water features remains relatively constant over this range of surface pressures.

By repeating this same analysis on the \texttt{Tiberius} spectrum, we measure chi-square values of $\chi^2 = 148.9$, $194.6$, and $137.9$ for the flat-line, 300 times solar, and 100\% H$_2$O models, respectively. This corresponds to Bayes factors of $\ln\mathcal{B} = 22.3$ for a flat line over the 300 times solar model, and of $\ln\mathcal{B} = 5.5$ for the 100\% H$_2$O model over a flat line, in agreement with the values obtained from the \texttt{exoTEDRF} spectrum.

The exact significance of this preference for a water feature is further quantified through the Bayesian evidence in section \ref{sec:significance}. We describe in the sections below the plausibility of this tentative water feature being stellar in nature as well as the constraints we can derive on any possible atmospheric water abundance. 

\subsection{Leveraging TOI-270\,d as a Stellar Contamination Control}

\begin{figure}
\includegraphics[width=0.48\textwidth]{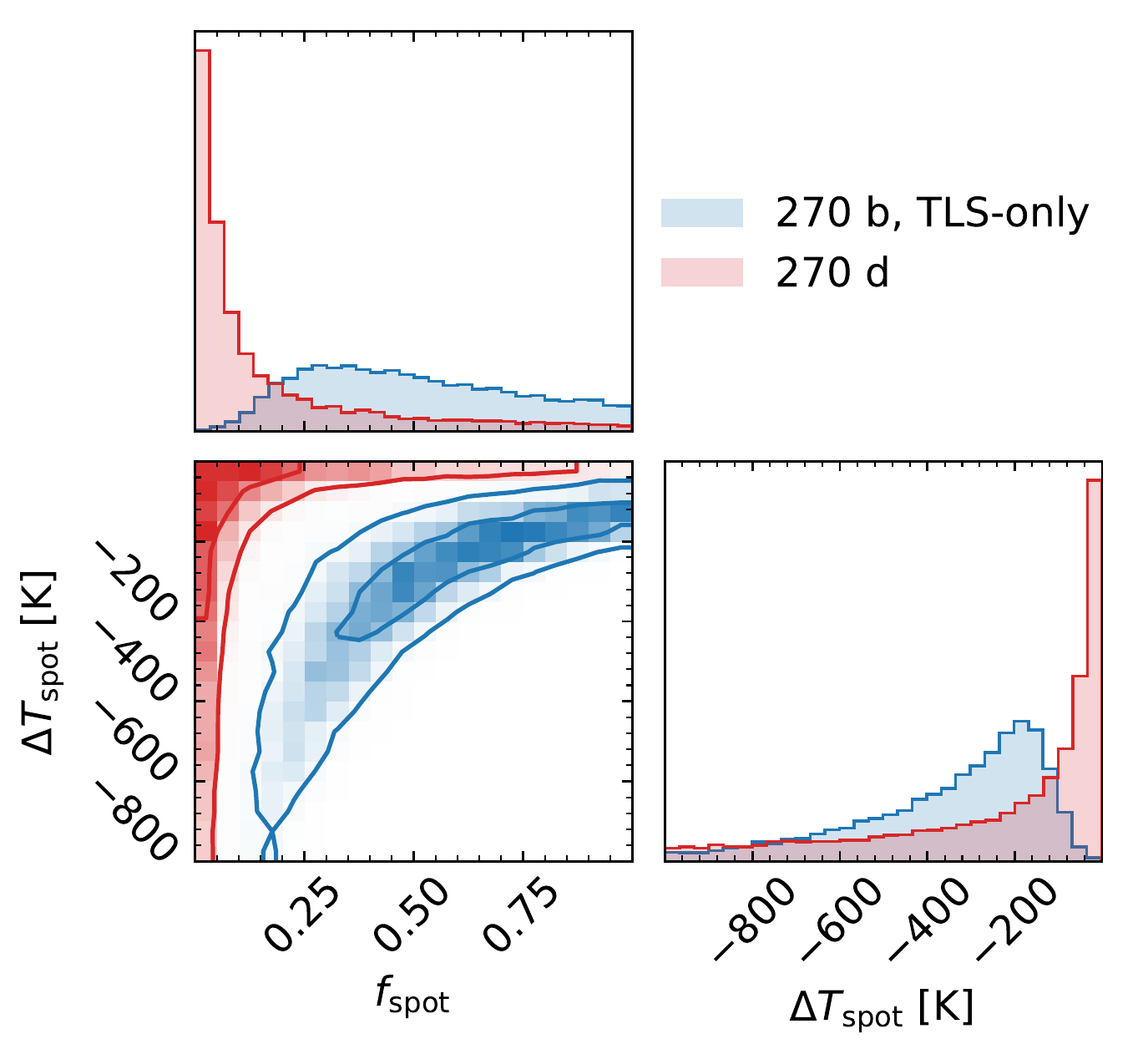}
\caption{Corner plot of the spot covering fraction and spot contrast from the bare-rock + free TLS retrieval on TOI-270\,b (blue), and the atmosphere + free TLS retrieval on TOI-270\,d (red). Owing to its larger size, the atmospheric retrieval on TOI-270\,d is much more constraining regarding the TLS parameters and is able to rule out significant contamination from the star. Moreover, the range of $f_\mathrm{spot}$ and $\Delta T_\mathrm{spot}$ values needed for TLS to explain TOI-270\,b's spectrum are significantly disfavored by the constraints from TOI-270\,d.}
\label{fig:tls_b_vs_d}
\end{figure}

When fitting the spectrum of TOI-270\,b accounting only for the transit light source effect and assuming no prior knowledge of the stellar contamination parameters, we find that the 2$\sigma$ confidence interval  of $f_\mathrm{spot}$ and $\Delta T_\mathrm{spot}$ covers (seemingly unphysical) combinations ranging from moderate spot covering fractions with large spot contrasts and vice versa (Figure \ref{fig:tls_b_vs_d}).
If we instead fit for the TLS parameters using the transmission spectrum of TOI-270\,d, we constrain $f_\mathrm{spot}$ and $\Delta T_\mathrm{spot}$ to be low -- consistent with 0 -- at high precision,
inconsistent with the range of values that would be needed to explain TOI-270\,b's spectrum from TLS alone.
It is possible that the stricter TLS constraints obtained from TOI-270\,d are not solely the result of its larger size, but also of the many molecular opacity bands observed over the NIRSpec/G395H wavelength range. These features provide a robust measurement of TOI-270\,d's continuum opacity, which may in turn make the atmospheric inferences less susceptible to degeneracies with the TLS effect.
The constraints from TOI-270\,d enable us to derive more stringent atmospheric inferences for TOI-270\,b while still marginalizing over the remaining uncertainty on the TLS parameters.

The lack of stellar contamination from TOI-270 is not unexpected given its low observed photometric modulation \citep{Gunther_2019}.
With a rotation period of $\sim$58 days \citep{Van_Eylen_2021}, magnetic activity is expected to be lower compared to M-stars that have rotation timescales of a few days \citep{Mohanty2003}, such as TRAPPIST-1 \citep[$\sim$3.3\,days,][]{Luger_2017} and GJ 3090 \citep[$\sim$18\,days,]{Almenara_2022,Ahrer_2025}. This statement is also supported by the observation of an H$\alpha$ absorption feature in the spectrum of TOI-270 \citep{Gunther_2019}, indicative of low magnetic activity. Lower activity levels would then result in smaller spot coverage and thus, in the lack of significant TLS effect in the observed spectra. This is consistent with the measured rotational modulation of TOI-270, on the order of 1\,ppt in the TESS bandpass \citep{Van_Eylen_2021}, corresponding to spot covering fractions of less than 2\% according to the relations of \citet{Rackham2018}. We note, however, that stellar contamination has been claimed as a potential explanation for some of the signal observed in the transmission spectra of rocky planets orbiting M dwarfs with rotation periods of more than 50 days \citep[e.g., GJ 486, GJ 1132, and LHS 1140;][]{Moran_2023,May2023,cadieux2024transmissionspectroscopyhabitablezone}, although these objects coincidentally have spectral types later than or equal to M3.5, at which point stars are expected to be fully convective \citep{Chabrier1997}. Given their lack of a tachocline -- the boundary between the radiative and convective regions \citep{Dikpati1999} -- these objects are possibly unable to sustain a solar-like dynamo \citep{Parker1955} and could be in a distinct activity regime.

\subsection{Free Chemistry Retrievals}

\begin{figure}
\begin{center}
\includegraphics[width=0.48\textwidth]{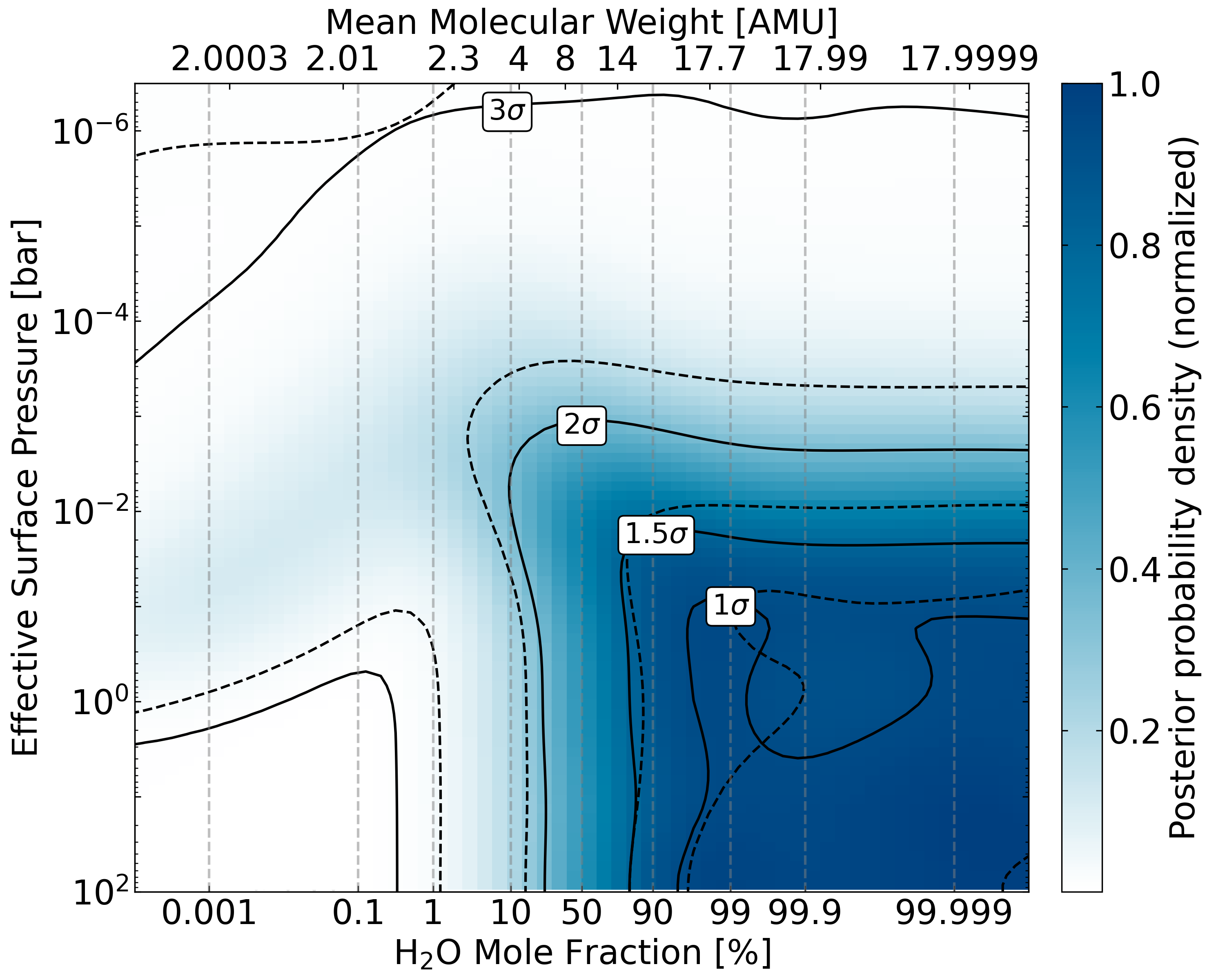}
\end{center}
\vspace{-5mm}\caption{Constraints on the water mole fraction and effective surface pressure from the free chemistry retrieval without TLS. Color represents the normalized posterior probability density from the retrieval assuming no TLS effect, and the full black lines indicate the 1, 1.5, 2, and 3$\sigma$ confidence intervals. The dashed lines represent those same confidence intervals for the retrieval where we assume as a prior the TLS parameter constraints from the retrieval on TOI-270\,d.} 
\label{fig:h2o_vs_pcloud}
\end{figure}

The H$_2$O/H$_2$ atmospheric retrieval performed without considering the transit light source effect shows 
a $\sim$2$\sigma$ confidence interval that encases water-rich ($>$50\% water) scenarios with effective surface pressures of more than 10$^{-3}$\,bar (Figure \ref{fig:h2o_vs_pcloud}). At the 2-3$\sigma$ level, H$_2$-rich atmosphere compositions are also allowed as the right combination of low MMW, water volume mixing ratio, and surface pressure can reproduce the 40\,ppm feature amplitude that would correspond to a water-rich scenario. While H$_2$-rich and H$_2$O-rich scenarios can in principle be distinguished from the width and relative amplitudes of the molecular features \citep{Benneke2013}, the partial water feature that is covered by NIRSpec/G395H and the precision of the observations are not sufficient to differentiate between these scenarios. Within the 3$\sigma$ level and beyond, scenarios with low effective surface pressures and thus, low water feature amplitudes, are not ruled out as a flat spectrum can reasonably explain the observed spectrum. Additionally, we rule out a portion of the parameter space at $>$3$\sigma$ confidence for water mole fractions below 0.3\% and 
effective surface pressures above 1\,bar, as the water feature amplitude produced by such scenarios is too great to reproduce the tentative feature.

\begin{figure*}
\begin{center}
\includegraphics[width=0.32\textwidth]{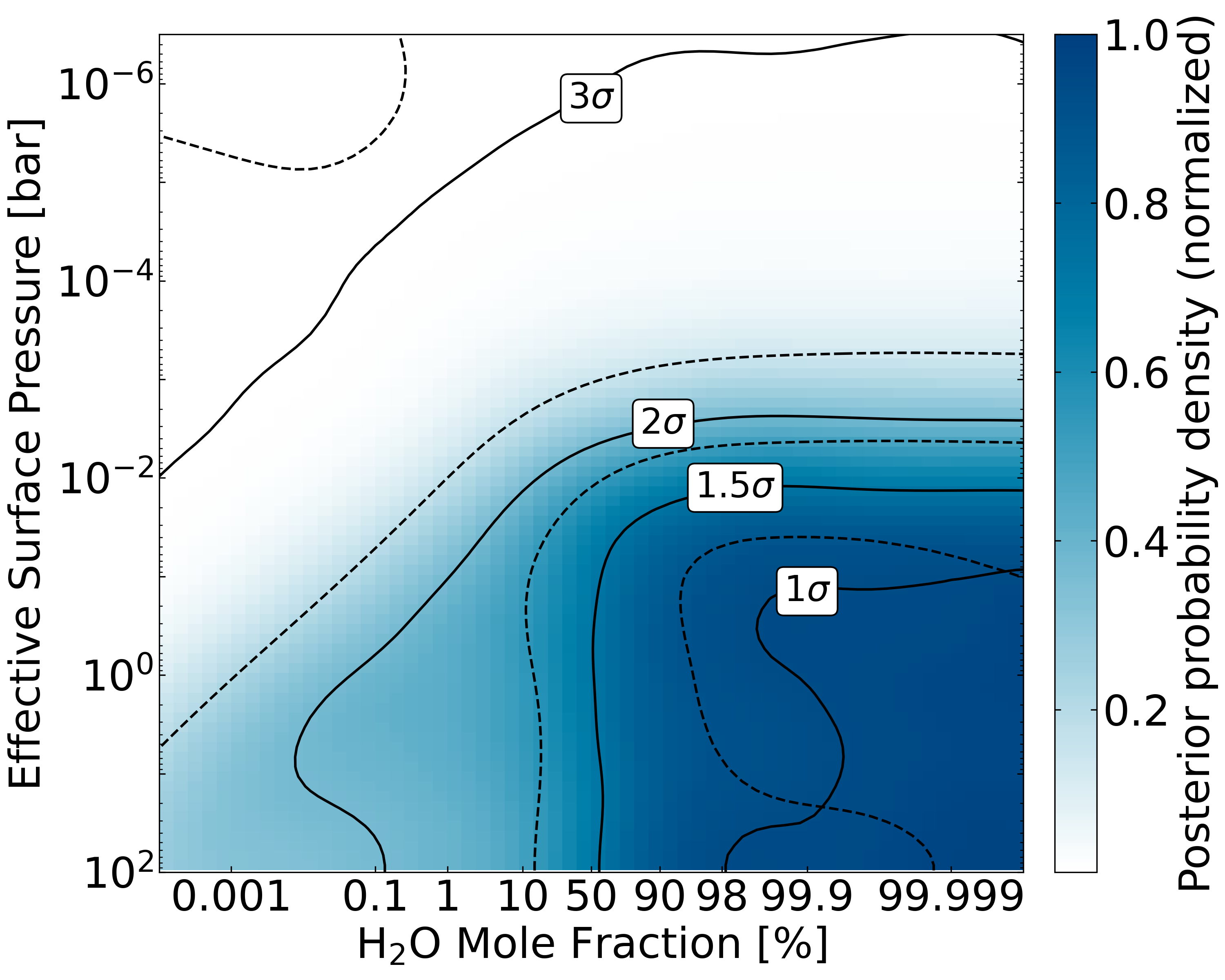}
\includegraphics[width=0.32\textwidth]{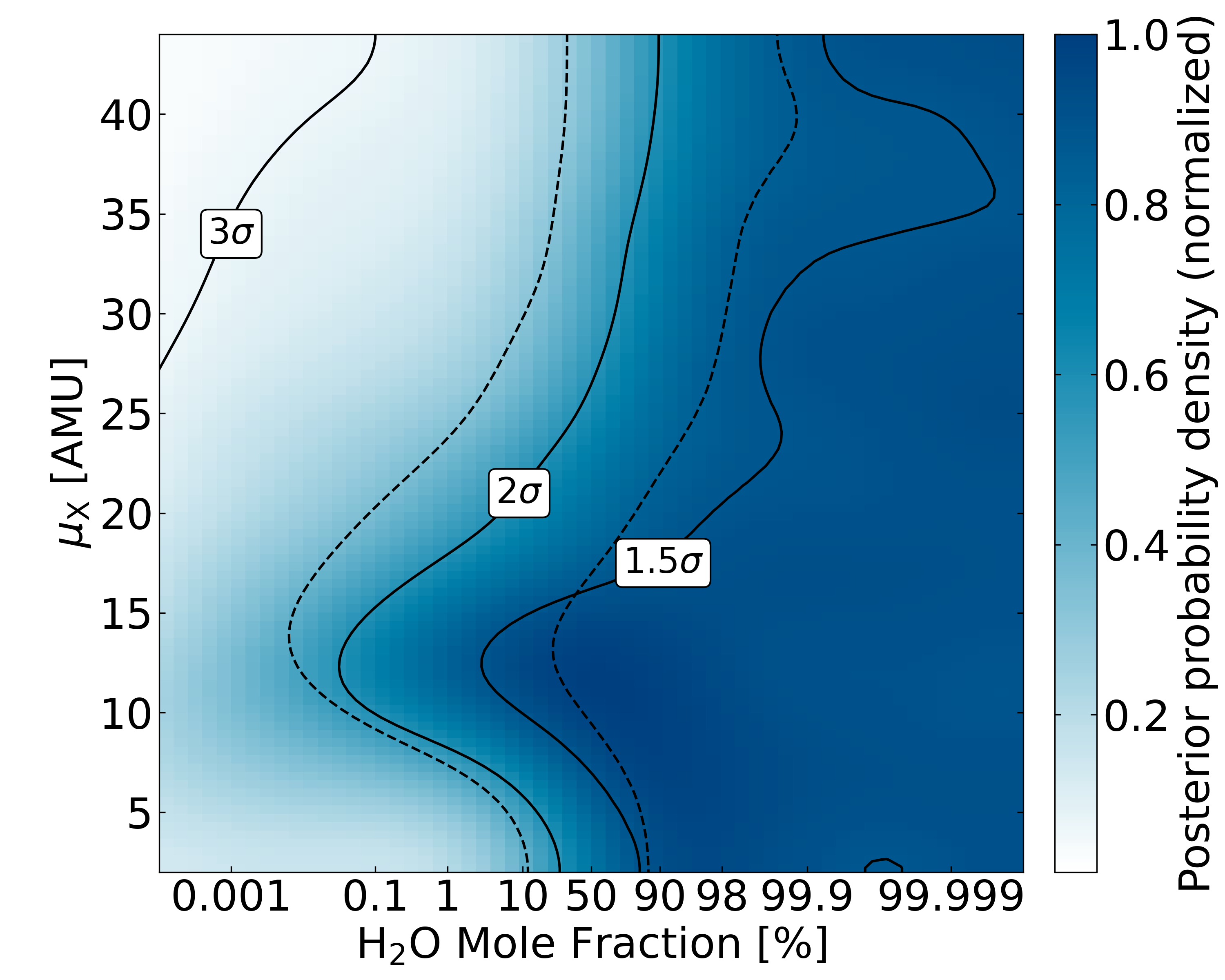}
\includegraphics[width=0.32\textwidth]{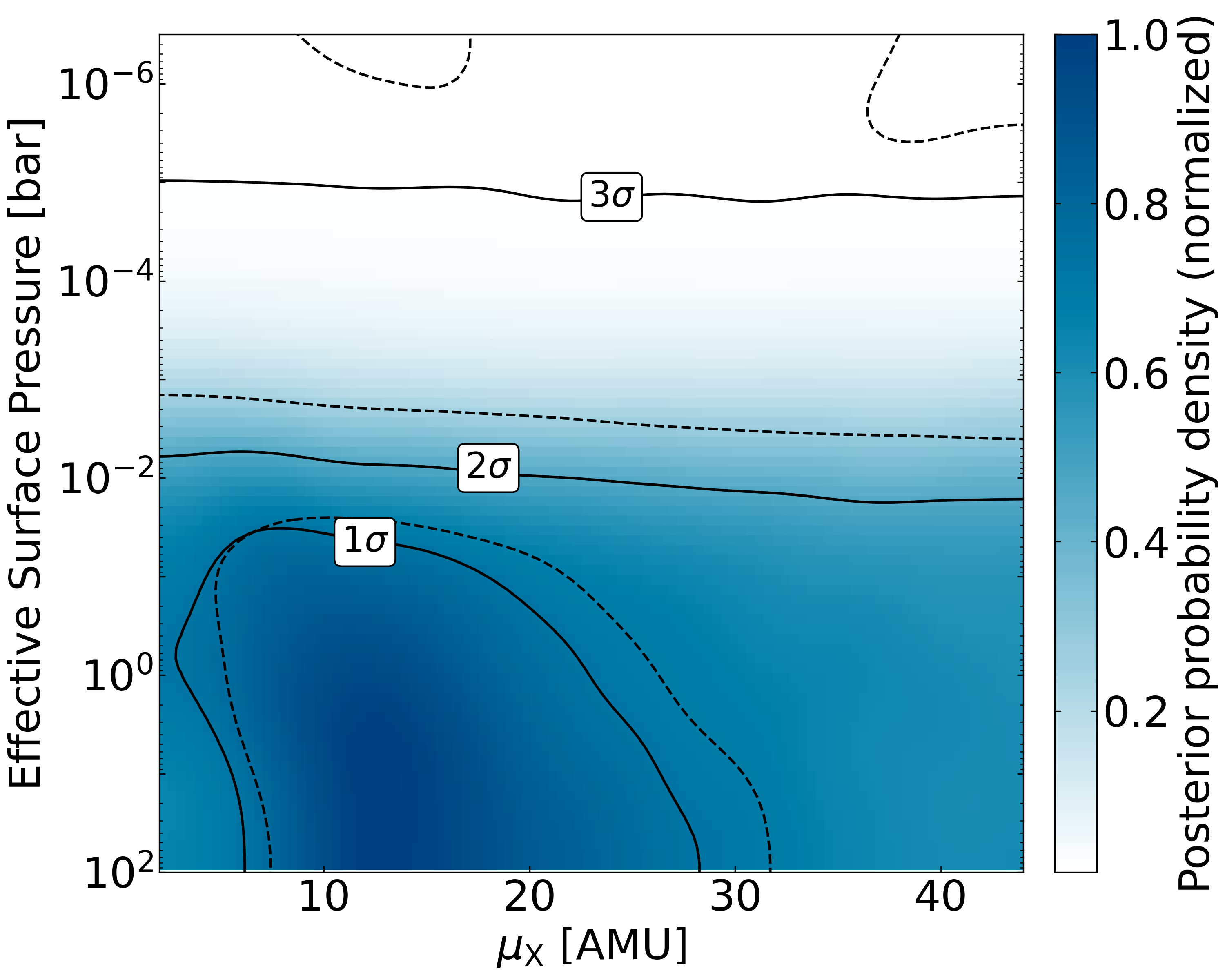}
\end{center}
\vspace{-5mm}\caption{Constraints from the atmospheric retrieval considering the fictitious species X as the background gas. \textbf{Left:} Joint constraints on the water mole fraction and effective surface pressure, where color represents the normalized probability density. The 1, 1.5, 2, and 3$\sigma$ probability contours are indicated by the full and dashed black lines for the retrievals assuming no TLS and with constrained TLS, respectively. \textbf{Middle:} Joint constraints on the water mole fraction and background gas MMW ($\mu_\mathrm{X}$). The 1.5, 2, and 3$\sigma$ probability contours are indicated by the full and dashed black lines for the retrievals assuming no TLS and with constrained TLS, respectively. Beyond water mole fractions of 90\%, the background MMW is unconstrained as the global MMW is dominated by that of H$_2$O. Below that, the 1.5 and 2$\sigma$ confidence regions are confined to a given range of MMW values, as global MMW that are too high/low would result in water absorption features that are too large/small compared to the tentative feature. \textbf{Right:} Joint constraints on the background gas MMW and effective surface pressure.} 
\label{fig:h2o_vs_pcloud_muX}
\end{figure*}

When including the MMW of the background gas as a free parameter in the atmospheric retrieval, we find that the 1.5$\sigma$ confidence region encases water mole fractions and effective surface pressures above 50\% and 10$^{-2}$\,bars (Figure \ref{fig:h2o_vs_pcloud_muX}), respectively, consistent with the results from the H$_2$O/H$_2$ retrieval. We find in this case, however, that scenarios with low water mole fractions and high surface pressures are allowed by the retrieval as the background MMW can adjust itself to conserve the correct feature amplitude. At the 2$\sigma$ confidence level, scenarios with water mole fractions below 10\% show background MMW $\mu_\mathrm{X}$ values of $\gtrsim$6\,AMU as the resulting water feature amplitude would otherwise be too large. Similarly, low H$_2$O VMR scenarios also disfavor high ($\lesssim18$\,AMU) $\mu_\mathrm{X}$ values as they would result in smaller feature amplitudes compared to a pure H$_2$O atmosphere.

When accounting for the TLS effect, we find that, given the strong preference for low spot temperature contrasts and spot covering fractions from the retrieval on TOI-270\,d, we obtain probability posterior distributions for the water mole fraction and cloud top pressure that are nearly identical in shape and precision compared to when TLS is not considered (Figs. \ref{fig:h2o_vs_pcloud} \& \ref{fig:h2o_vs_pcloud_muX}). This further confirms that the level of stellar contamination that is allowed by the constrained TLS retrieval is not sufficient to produce the tentative water absorption signature near 3\,$\mu$m, which remains best explained by water-rich atmosphere scenarios.

\subsection{Offset Between the NRS1 and NRS2 Detectors}\label{sec:offset}

When including an offset between the two detectors for the retrievals performed on the \texttt{exoTEDRF} reduction, we measure $\delta_\mathrm{NRS1,2}$ values that are 2--2.5$\sigma$ away from 0. The magnitude of the offsets are similar despite differences in modeling assumptions, with values of $26\pm11$\,ppm, $26\pm10$\,ppm, and $20\pm12$\,ppm for the H$_2$/H$_2$O, X/H$_2$O, and free TLS retrievals, respectively. However, when including an offset in the retrieval on TOI-270\,d, we measure a value that is consistent with 0 ($\delta_\mathrm{NRS1,2} = 2\pm18$\,ppm). One possibility is thus that the non-zero offsets measured in the TOI-270\,b retrievals result from missing opacity sources rather than from a systematic offset between the NRS detectors.

When an offset is included in the retrievals, we find that the overall shape of the H$_2$O mole fraction vs. effective surface pressure are relatively unchanged (Figures \ref{fig:h2o_vs_pcloud_offset} and \ref{fig:h2o_vs_pcloud_muX_offset}), although the constraints are overall weaker as the offset is able to partially mute the amplitude of the feature attributed to water.

\subsection{Statistical Significance}\label{sec:significance}

We compare the Bayesian evidences from our atmosphere-only retrievals to that of a simple “flat line” model to quantify the significance of the preference for an atmosphere model. The Bayesian evidence for the flat line model is computed by running a retrieval fitting for the abundance of He (spectrally inactive) with X as the fill gas, which is then compared to a retrieval where we fit for both the abundances of He and H$_2$O (again using X as the fill gas). We use this specific setup as Bayesian model comparison requires that the models be nested \citep{Trotta_2008}.
When including H$_2$O in the retrievals, we find an increase in the evidence corresponding to Bayes factors of $\ln\mathcal{B}$ = 0.8 (inconclusive) when including an offset and 3.2 (moderate) when no offset is considered. Similarly, when repeating these retrievals on the \texttt{Tiberius} spectrum, we measure Bayes factors of $\ln\mathcal{B}$ = 0.3 and 2.9 with and without offset, respectively, indicating that the significance of the preference for a water absorption feature depends heavily on the inclusion of an offset parameter in the retrievals. 

While it is shown in section \ref{sec:SCmods} and Fig. \ref{fig:spec_vs_mods} that a water-rich atmosphere provides a better fit to the data (lower $\chi^2$) than a flat-line model, it does not necessarily provide a good fit in an absolute sense. Indeed, the reduced chi-square of the 100\,bar pure H$_2$O model to the \texttt{exoTEDRF} spectrum is of $\chi^2/N = 1.40$. Similarly, the X+H$_2$O retrievals performed on the \texttt{exoTEDRF} spectrum result in $\chi^2/N$ values of 1.39 and 1.34 with and without an offset, respectively. With its slightly larger uncertainties, the \texttt{Tiberius} spectrum produces $\chi^2/N$ values of 1.21 (no offset) and 1.13 (with offset), closer to the value of 1 that would be expected for independent and Gaussian-distributed residuals. Possible explanations for these elevated chi-square values include an underestimation of the transit depth uncertainties, as well as residual astrophysical and instrumental signal in the spectrum that is not fitted by the model.

\section{Atmospheric Evolution Analysis}\label{sec:evolution}

\subsection{Coupled Geochemical Evolution Modeling}

To investigate the plausibility of different atmospheric scenarios, we use the PACMAN-P geochemical evolution model to simulate the coupled atmosphere-interior evolution of TOI-270\,b \citep{krissansen2024erosion}. The framework has previously been applied to model the transition from primary to secondary atmospheres, and can accommodate arbitrary C, H, and O-bearing atmospheric compositions in contact with a molten interior. PACMAN-P simultaneously solves for geochemical and thermal equilibrium as the magma ocean solidifies. At each time step, we calculate the multiphase equilibrium of all volatiles and Fe-bearing species between the atmosphere and the molten silicate interior. The resulting atmospheric species determine surface temperature, as calculated using a radiative-convective climate model that balances absorbed stellar radiation, outgoing longwave radiation, and internal heatflow. Internal heatflow is determined by parameterized mantle convection with temperature-dependent mantle viscosity, and driven by the heat of accretion, assumed radionuclides, and the latent heat of mantle solidification. Imposed stellar bolometric and XUV evolution drives atmospheric escape, which is either XUV-limited or diffusion limited, depending on the composition of the upper atmosphere (H, C, and O-bearing species can escape) \citep{krissansen2024erosion}.

\subsection{Possible Evolution Scenarios of TOI-270\,b}

\begin{figure}
\begin{center}
\includegraphics[width=0.48\textwidth]{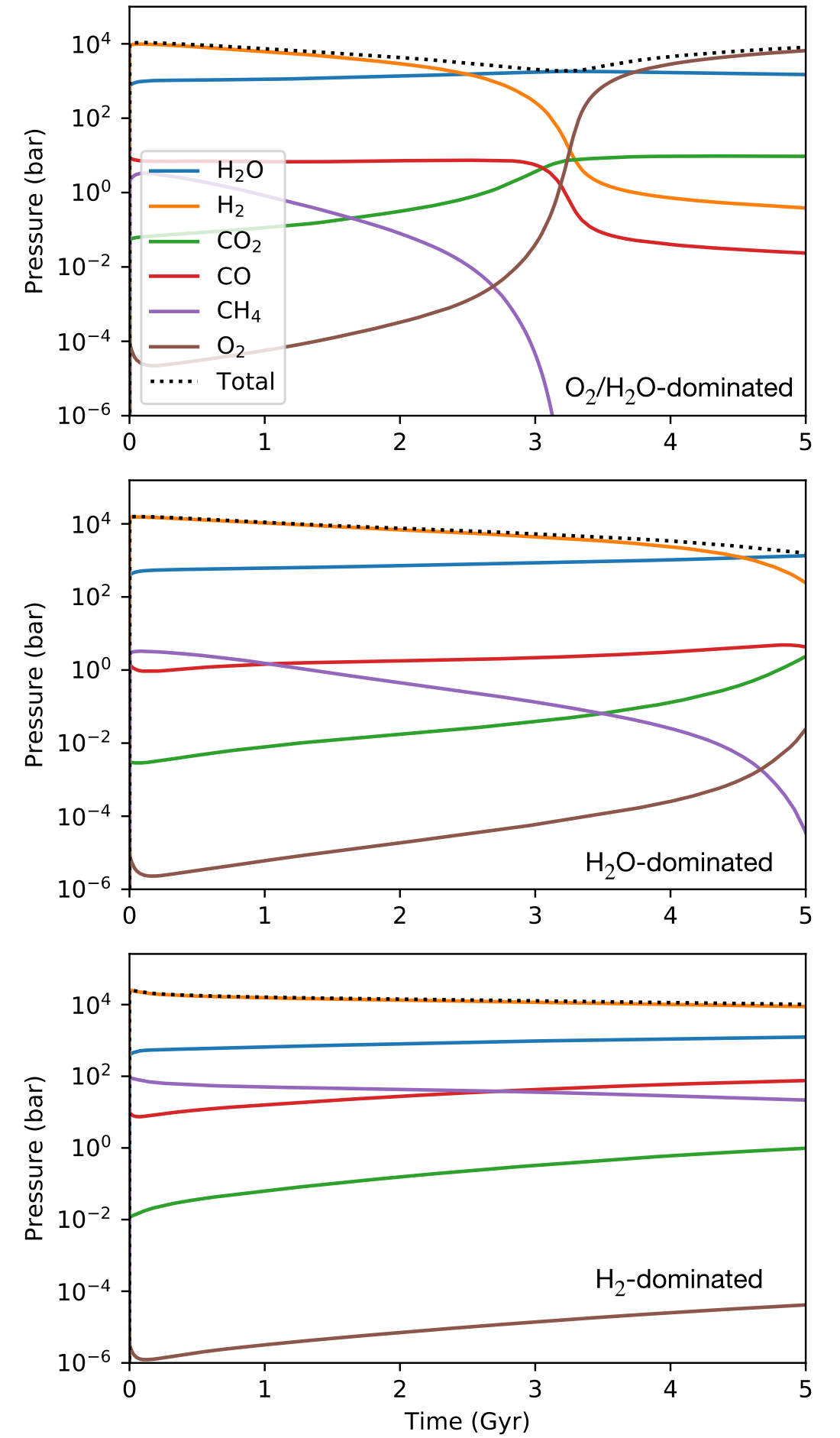}
\end{center}
\vspace{-5mm}\caption{Coupled atmosphere-interior evolution modelling shows that a diversity of atmospheric outcomes are consistent with observational constraints. The PACMAN-P geochemical evolution model \citep{krissansen2024erosion} was used to explore plausible evolutionary scenarios for TOI-270\,b. This model explicitly couples magma ocean solidification, speciation of C, H, O, and Fe-bearing species, stellar evolution, thermal escape, and radiative-convective climate. Subpanels represent scenarios whereby hydrogen escape causes a transition to an O$_2$-H$_2$O atmosphere (top), an H$_2$O-dominated atmosphere (middle), and a case where an H$_2$-dominated atmosphere is retained (bottom). The internal volatile reservoirs of the planet are not shown.} 
\label{fig:Three_Scenarios_TOI_270b}
\end{figure}

Three illustrative outputs from the atmosphere-interior evolution modelling of TOI-270\,b are shown in Figure \ref{fig:Three_Scenarios_TOI_270b}. These results are drawn from a larger Monte Carlo ensemble that explores sensitivity to initial volatile inventories, and uncertainties in stellar evolution, atmospheric escape, and other planetary parameters \citep{krissansen2024erosion}. We find that a diversity of atmospheric outcomes can be reconciled with the current observational constraints, and three qualitatively different examples of this are highlighted in Figure \ref{fig:Three_Scenarios_TOI_270b}. The top panel of Fig.~\ref{fig:Three_Scenarios_TOI_270b} shows a scenario whereby TOI-270\,b is initialized with a comparatively oxidized volatile inventory (1.1x10$^{23}$\,kg H$_2$O, 5.5x10$^{22}$ kg H$_2$, and 2.1x10$^{20}$ kg C). Net hydrogen escape causes a transition to a strongly oxidizing atmosphere  (approximately 20\% H$_2$O and 80\% O$_2$ by volume), which is broadly compatible with the fictitious background constraints in Fig.~\ref{fig:h2o_vs_pcloud_muX} ($\mu_\mathrm{O_2}$ = 32 AMU). The middle panel of Fig.~\ref{fig:Three_Scenarios_TOI_270b} shows an intermediate scenario whereby the planet is initialized with a more reducing volatile composition (3.3x10$^{21}$ kg H$_2$O, 9.8x10$^{22}$ kg H$_2$, and 1.1x10$^{20}$ kg C). Here, hydrogen loss causes a transition to a steam-dominated atmosphere (approximately 85\% H$_2$O and 15\% H$_2$ by volume), which is within the 1.5$\sigma$ confidence region of the H$_2$/H$_2$O atmospheric retrieval (Fig.~\ref{fig:h2o_vs_pcloud}). Finally, the lower panel of Fig.~\ref{fig:Three_Scenarios_TOI_270b} shows an example of a similarly reduced scenario but with a larger initial volatile inventory that is more resistant to escape (6.2x10$^{21}$ kg H$_2$O, 1.6x10$^{23}$ kg H$_2$, and 2.7x10$^{21}$ kg C). Here, TOI-270\,b has retained a H$_2$-dominated atmosphere  (approximately 85\% H$_2$ and 15\% H$_2$O by volume). This is consistent within 2$\sigma$ with the results from the H$_2$/H$_2$O retrieval.
In all three scenarios, the dense overlying atmosphere maintains a deep magma ocean throughout, within which substantial dissolved volatiles are stored, and gradually exsolved to offset escape losses.

\section{Discussion}\label{sec:discussion}

The relatively low density of TOI-270\,b and possible evidence for a water absorption feature in its spectrum both hint at the presence of a significant amount of volatiles on TOI-270\,b, consistent with expectations for a “water world” scenario. This type of planet, with volatile inventories much higher than the Solar System terrestrial planets, has been predicted by formation theory \citep[e.g.,][]{Lichtenberg_2019,Venturini2020,Izidoro_2022,burn2024radiusvalleymigratedsteam}. Moreover, planet formation modeling studies have found that water-rich formation is favored for M dwarf star systems \citep{Burn_2021,Venturini_2024}. Mass-radius measurements of planets have revealed a few “water world” candidates with densities consistent with significant water mass fractions \citep[e.g.,][]{Luque_2022,Piaulet_2022,cadieux2023newmassradiusconstraints}, and recent HST/JWST observations have provided the first spectroscopic evidence for their existence \citep{Roy_2023,Piaulet_Ghorayeb_2024}. We emphasize that the measurements presented in this work are only tentative, and any convincing evidence of an atmosphere on a rocky exoplanet should ideally come from repeatable measurements of atomic or molecular features. Nevertheless, we shortly discuss below the implications that the discovery of volatiles on TOI-270\,b, if confirmed, would signify in terms of its nature, formation, and evolution.

\subsection{Formation of TOI-270\,b}

If TOI-270\,b formed beyond the water snowline, it could have accreted a significant amount of volatiles throughout its formation as water would be directly available in the solid phase \citep{_berg_2011}. Considering the effective temperature of TOI-270 \citep[3506\,K;][]{Van_Eylen_2021} and assuming a water evaporation temperature of 135\,K \citep{_berg_2011}, the position of the water snowline would be at 0.6\,AU, 20 times further out than the current-day orbital position of TOI-270\,b (Table \ref{table:wlc_fit}). We note, however, that the snowlines would have most likely been further out at the time of planet formation, as the star would have been brighter during its pre-main sequence phase \citep{Baraffe2015}. Such a scenario would require significant inward migration after the planet has formed, which is consistent with the fact that the TOI-270 planets are in near-resonances, indicative of a potential migration history for the system \citep{Melita1996,Lee2002,Terquem2007}. This is akin to the formation history of the Galilean moons of Jupiter, where Ganymede and Callisto formed beyond the snowline of Jupiter's circumplanetary disk, amassed $\sim$50\% water by mass, and subsequently migrated inward \citep{Pollack1974,Sasaki2010}.


Another possibility is that, rather than the volatile inventory of TOI-270\,b resulting from the direct accretion of icy material from the protoplanetary disk, it is the result of interaction between a primordial H$_2$/He atmosphere with the interior of the planet \citep{Dorn_2021,schlichting2022chemicalequilibriumcoresmantles,Young_2023}. The interaction of hydrogen with the mantle produces water through chemical reactions such as FeO$_\mathrm{(l)}$ + H$_2$$_\mathrm{(g)}$ → H$_2$O$_\mathrm{(g)}$ + Fe$_\mathrm{(l)}$ \citep{Kite2020,Kite_2021}. The primordial atmosphere would then gradually be lost to space after dissipation of the protoplanetary disk, leaving behind a rocky planet enriched in volatiles (Fig. \ref{fig:Three_Scenarios_TOI_270b} top and middle panels). Simulations presented in \citet{rogers2024fleetingforgottenimprintescaping} show that 0.5--1.0\% of water-by-mass can be produced through this process, within the same order of magnitude as our inferred WMF as well as the volatile mass fraction measured on TOI-270\,d \citep[6.0$^{+3.7}_{-2.3}\%$, 1244$^{+761}_{-464}$\,M$_{\oplus,\mathrm{ocean}}$;][]{benneke2024jwstrevealsch4co2}.
Surface volatiles can then be continually replenished through equilibrium exsolution from the magma ocean.
The total planetary volatile inventories presented in Fig. \ref{fig:Three_Scenarios_TOI_270b} are broadly consistent with the mass constraints of the planet.

\subsection{TOI-270\,b in the Context of the Cosmic Shoreline}

\begin{figure}[h]
\begin{center}
\includegraphics[width=0.5\textwidth]{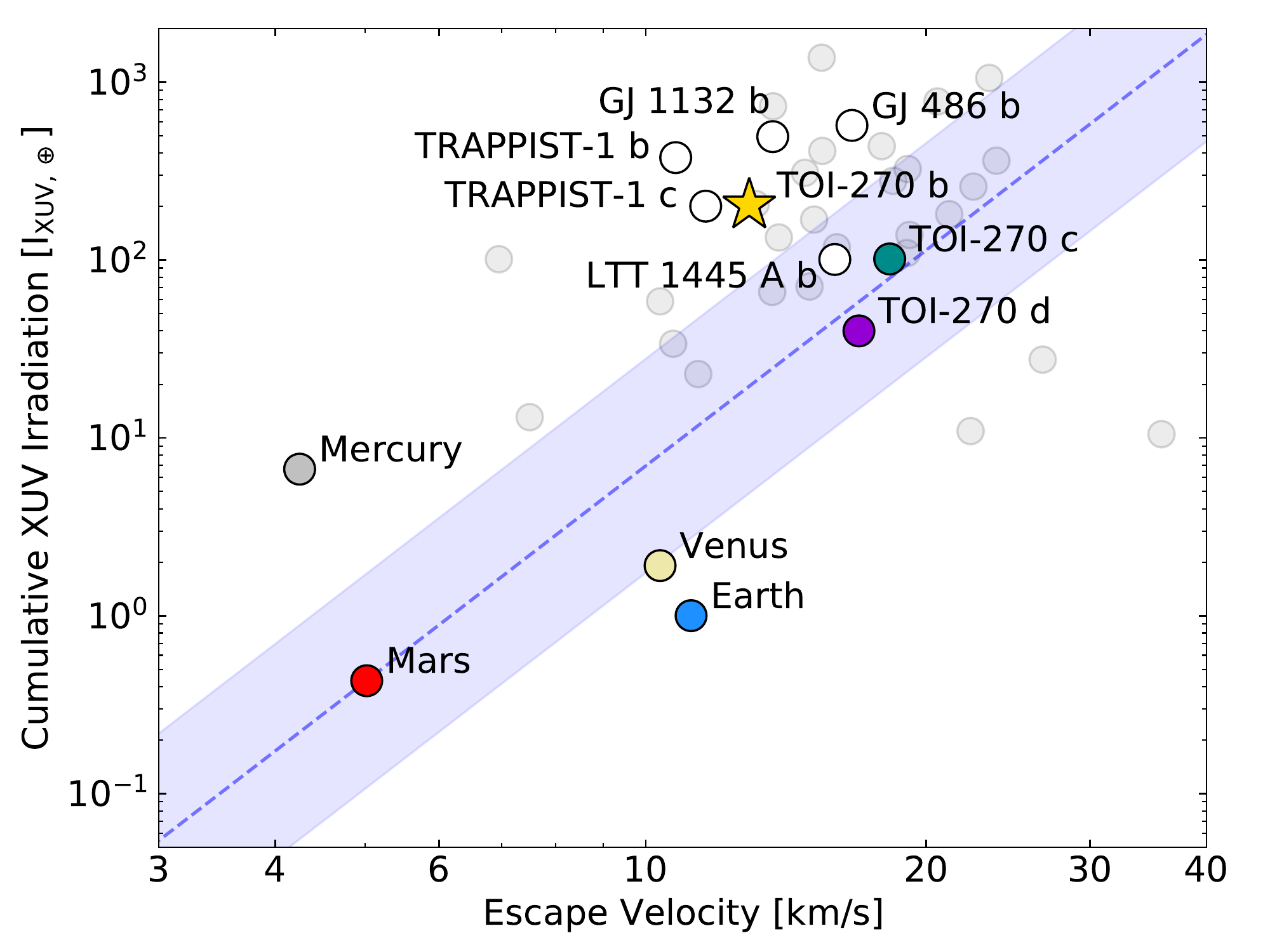}
\end{center}
\vspace{-5mm}\caption{Position of TOI-270\,b relative to the cosmic shoreline. Escape velocity and cumulative XUV irradiation, computed using the scaling of \citet{Zahnle_2017}, for all known exoplanets with equilibrium temperatures below 1,000\,K, radii smaller than 1.8\,$R_\oplus$, and masses measured to higher than 20\% precision \citep[data taken from the NASA Exoplanet Archive,][]{Akeson2013}. The blue dashed line indicates the position of the cosmic shoreline shown in \citet{Zahnle_2017}, and we add a shaded region corresponding to a factor of four uncertainty in XUV irradiation, as in \citet{xue2024jwstthermalemissionterrestrial}. The solar system telluric planets are shown for comparison, and we also indicate the position of TRAPPIST-1\,b \citep{Greene_2023}, TRAPPIST-1\,c \citep{Zieba_2023}, GJ 1132\,b \citep{xue2024jwstthermalemissionterrestrial}, GJ 486\,b \citep{mansfield2024thickatmosphereterrestrialexoplanet}, and LTT 1445\,Ab \citep{wachiraphan2024thermalemissionspectrumnearby}, which have all been found to be consistent with a dark, bare-rock scenario from JWST thermal emission observations. However, some of these targets, such as TRAPPIST-1\,c \citep{Zieba_2023} remain consistent with a variety of atmospheric scenarios.
The two sub-Neptunes of the TOI-270 system, planets c and d, are also shown for comparison with planet b.} 
\label{fig:Vesc_vs_Irr}
\end{figure}

The mass, radius, and instellation of TOI-270\,b place it close to the cosmic shoreline \citep[][Figure \ref{fig:Vesc_vs_Irr}]{Zahnle_2017}, on the “airless” side, in the same region as planets shown to be consistent with bare rocks from JWST/MIRI thermal emission observations. If we also compare the positions of TOI-270\,c and d relative to the cosmic shoreline, they are closer to the no-atmosphere regime than Venus and Earth, with TOI-270\,c being at virtually the same distance from the shoreline as Mars. However, there is no doubt that TOI-270\,c and d host atmospheres based on their measured densities alone \citep[$\rho_\mathrm{p,c}$ = 2.7 and $\rho_\mathrm{p,d}$ = 2.9 g/cm$^3$;][]{Kaye2022}. One possible explanation is that the cumulative XUV irradiation received by the TOI-270 planets is overestimated by the scaling relation used in \citet{Zahnle_2017}, possibly to the point where TOI-270\,b could also sustain an atmosphere. 

More broadly, it is important to highlight that the cosmic shoreline is a quasi-empirical relationship with a functional form motivated by atmospheric escape theory. While the cosmic shoreline may separate airless worlds and planets with atmospheres at the population level, for any individual planet, atmospheric retention is a function of inital volatile inventories. If TOI-270\,b does have an atmosphere, it was possibly shielded from complete loss due to the large amount of volatiles it might have accreted. For example, Fig. \ref{fig:Three_Scenarios_TOI_270b} (bottom) shows a plausible evolutionary scenario whereby TOI-270\,b retains a hydrogen-rich atmosphere due to possessing a large initial volatile inventory and storing volatiles in the magma ocean, which never solidifies for all three cases shown. A consequence of this is that, rather than focusing on irradiation alone, there is great interest in determining the conditions that fare most favorable to the accumulation of volatiles at the time of and after the formation of a rocky planet in the search for secondary atmospheres.

\section{Conclusion}\label{sec:conclusion}

In this work, we present an updated density measurement and transmission spectrum for TOI-270\,b from its NIRSpec/G395H transit that was captured during a visit of the outermost planet of the system, the sub-Neptune TOI-270\,d. We find its density to be inconsistent at 4.4$\sigma$ with an Earth-like composition. It is instead consistent at 2.4$\sigma$ with a pure rock composition, which is unlikely if we assume the stellar iron-to-magnesium abundance ratio (Fe/Mg = $0.81\pm0.22$) to be representative of the planet's core mass fraction of the interior, unless it formed as a sub-Neptune in which case its hydrogen-rich past is expected to result in under-dense cores \citep{schlichting2022chemicalequilibriumcoresmantles}. Furthermore, we find the planet is best explained by non-zero, \%-level water by mass 
when assuming the stellar and solar priors for the core mass fraction, respectively. By comparing the transmission of TOI-270\,b spectrum to self-consistent models, we find that the data rule-out clear, solar-C/O atmospheres with metallicities $\leq$300 times solar. Moreover, atmospheric retrievals show a preference for the presence of a partial water absorption band at the bluest wavelengths of the NRS1 detector, whose amplitude is consistent with expectations for a water-rich atmosphere. However, planetary water absorption features are generally degenerate with the transit light source effect for M-dwarf host stars. To rule out this possibility, we leverage the transit of TOI-270\,d observed within the same visit, which is 3 times more sensitive to TLS given its larger size, and show that there is no evidence of significant stellar contamination in the spectrum. Finally, we find that the significance of the preference for the inclusion of water, which ranges from inconclusive to moderate, depends heavily on whether an offset between the two NRS detectors is considered in the retrievals and, to a lesser extent, data reduction methodologies.

Observations of TOI-270\,b in thermal emission using the 15\,$\mu$m photometric filter of the MIRI instrument, taken as part of the Hot Rocks Survey (GO 3730, P.I.: H. Diamond-Lowe), will provide important constraints about its heat redistribution efficiency and Bond albedo. More specifically, thermal emission measurements are highly sensitive to the overall extent of the possible atmosphere and will prove complementary with the transmission measurements, which are consistent with a wide range of surface pressures ($P=10^2-10^{-3}$\,bar within 2$\sigma$; Figure \ref{fig:h2o_vs_pcloud_muX}). Nevertheless, photometric secondary eclipse observations are subject to degeneracies depending on the minerological makeup of their surface as well as the composition and thermal struture of their possible atmosphere \citep[e.g.,][]{hammond2024reliabledetectionsatmospheresrocky,alam2024jwstcompassnearmidinfrared,Ducrot_2024}, and further transmission spectroscopy observations of TOI-270\,b are needed to confirm or rule-out the presence of a significant secondary atmosphere. The near-IR wavelengths covered by JWST NIRISS/SOSS (\citealt{doyon2023nearinfraredimagerslitless,Albert_2023}, $\lambda$ = 0.6--2.85\,$\mu$m) are ideal for such a task as they cover multiple water absorption features \citep[e.g.,][]{Feinstein_2023}. Near-simultaneous transit observations of multiple planets within a single system, such as the one presented in this work, could prove a powerful tool to mitigate stellar contamination in the search for secondary atmospheres on rocky planets outside our Solar System.

\vspace{10mm}
\section*{Acknowledgments}
We thank the anonymous reviewer for their insightful comments that improved the quality of this work. This work is based on observations with the NASA/ESA/CSA James Webb Space Telescope, obtained at the Space Telescope Science Institute (STScI) operated by AURA, Inc. All of the data presented in this paper were obtained from the Mikulski Archive for Space Telescopes (MAST) at the Space Telescope Science Institute. The specific observations analyzed can be accessed via \dataset[doi: 10.17909/qedm-gv97]{https://doi.org/10.17909/qedm-gv97}. L.-P.C. and B.B. acknowledge financial support from the Canadian Space Agency under grant 23JWGO2A05. L.-P.C. would like to thank Elsa Ducrot for helpful feedback on the manuscript. A.L'H. acknowledges support from the FRQNT under file \#349961. C.P.-G. acknowledges support from the E. Margaret Burbidge Prize Postdoctoral Fellowship from the Brinson Foundation. Y.M. acknowledges support from the European Research Council (ERC) under the European Union’s Horizon 2020 research and innovation programme (grant agreement no. 101088557, N-GINE). H.E.S and H.A. acknowledge support from STScI under grant JWST-GO-04098.005-A. R.A. acknowledges the Swiss National Science Foundation (SNSF) support under the Post-Doc Mobility grant P500PT\_222212 and the support of the Institut Trottier de Recherche sur les Exoplanètes (iREx). J.P.W. acknowledges support from the Canadian Space Agency (CSA) [24JWGO3A-03] and the Trottier Family Foundation via the Trottier Postdoctoral Fellowship held at the Institute for Research on Exoplanets (IREx). Research activities of the Board of Observational and Instrumental Astronomy at the Federal University of Rio Grande do Norte are supported by continuous grants from the Brazilian funding agencies CNPq. This study was financed in part by the CAPES-Print program. B.L.C.M. acknowledge CNPq research fellowships grant No. 305804/2022-7. We thank the Swiss National Science Foundation (SNSF) and the Geneva University for their continuous support of our planet search programs. This work has been in particular carried out in the frame of the National Centre for Competence in Research PlanetS supported by the SNSF. A.S.M. acknowledges financial support from the Spanish Ministry of Science and Innovation (MICINN) project PID2020-117493GB-I00 and from the Government of the Canary Islands project ProID2020010129.

\bigskip 
\noindent
\textit{Software:} \texttt{Astropy} \citep{astropy:2013,astropy:2018,astropy:2022}, \texttt{batman} \citep{Kreidberg_2015}, \texttt{emcee} \citep{Foreman_Mackey_2013}, \texttt{Matplotlib} \citep{Hunter:2007}, \texttt{NumPy} \citep{harris2020array}, and \texttt{SciPy} \citep{2020SciPy-NMeth}.

\clearpage

\appendix

\begin{figure*}[h]
\begin{center}
\includegraphics[width=1.\textwidth]{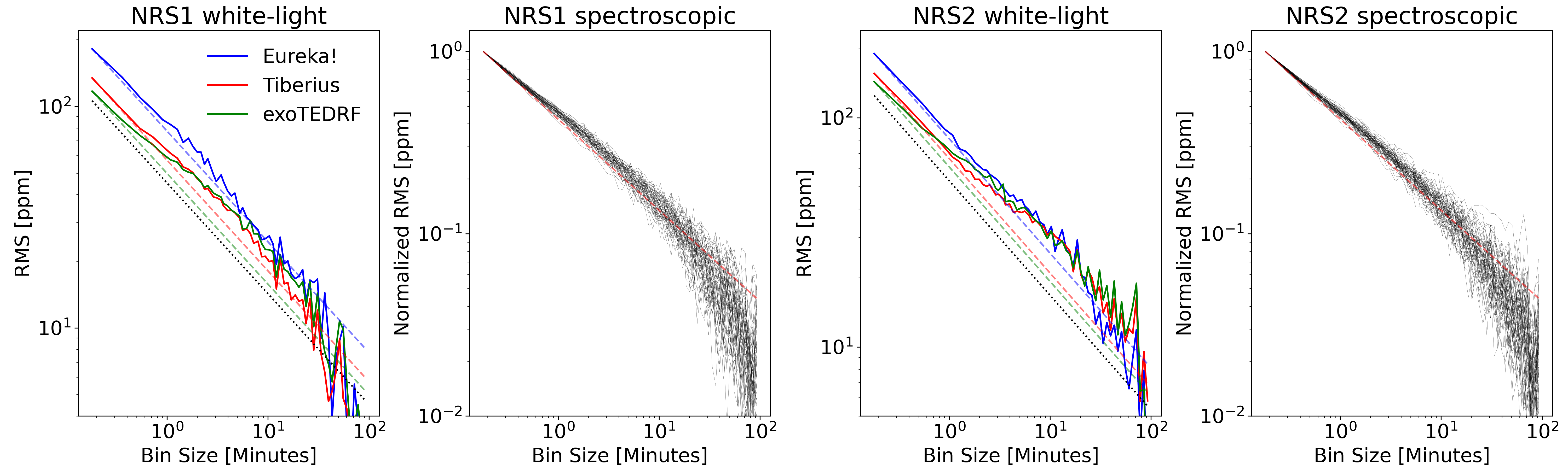}
\end{center}
\vspace{-5mm}\caption{RMS of the residuals as a function of bin size of the white and spectroscopic light curves. \textbf{Left:} RMS curves computed from the residuals of the fits to the NRS1 white light curves extracted using \texttt{Eureka!} (blue), \texttt{Tiberius} (red), and \texttt{exoTEDRF} (green). The colored dashed lines represent the expected 1/$\sqrt{N}$ trend of Poisson noise considering the fitted light-curve scatter. The black dashed line corresponds to the Poisson noise trend assuming the errors returned by the \texttt{jwst} pipeline. The three reductions show different scatters at the temporal resolution of the data but reach the same noise floor for bin sizes of $>$30--40\,minutes. \textbf{Center Left:} Normalized RMS curves (black lines) for all \texttt{exoTEDRF} spectroscopic light curves from the NRS1 detector. The red dashed line corresponds to the Poisson noise trend. \textbf{Center Right:} Same as \textbf{Left} for the NRS2 white light curves from the three independent reductions. \textbf{Right:}  Normalized RMS curves for all \texttt{exoTEDRF} spectroscopic light curves from the NRS2 detector.} 
\label{fig:RMS_plots}
\end{figure*}

\begin{figure*}[h]
\begin{center}
\includegraphics[width=1.\textwidth]{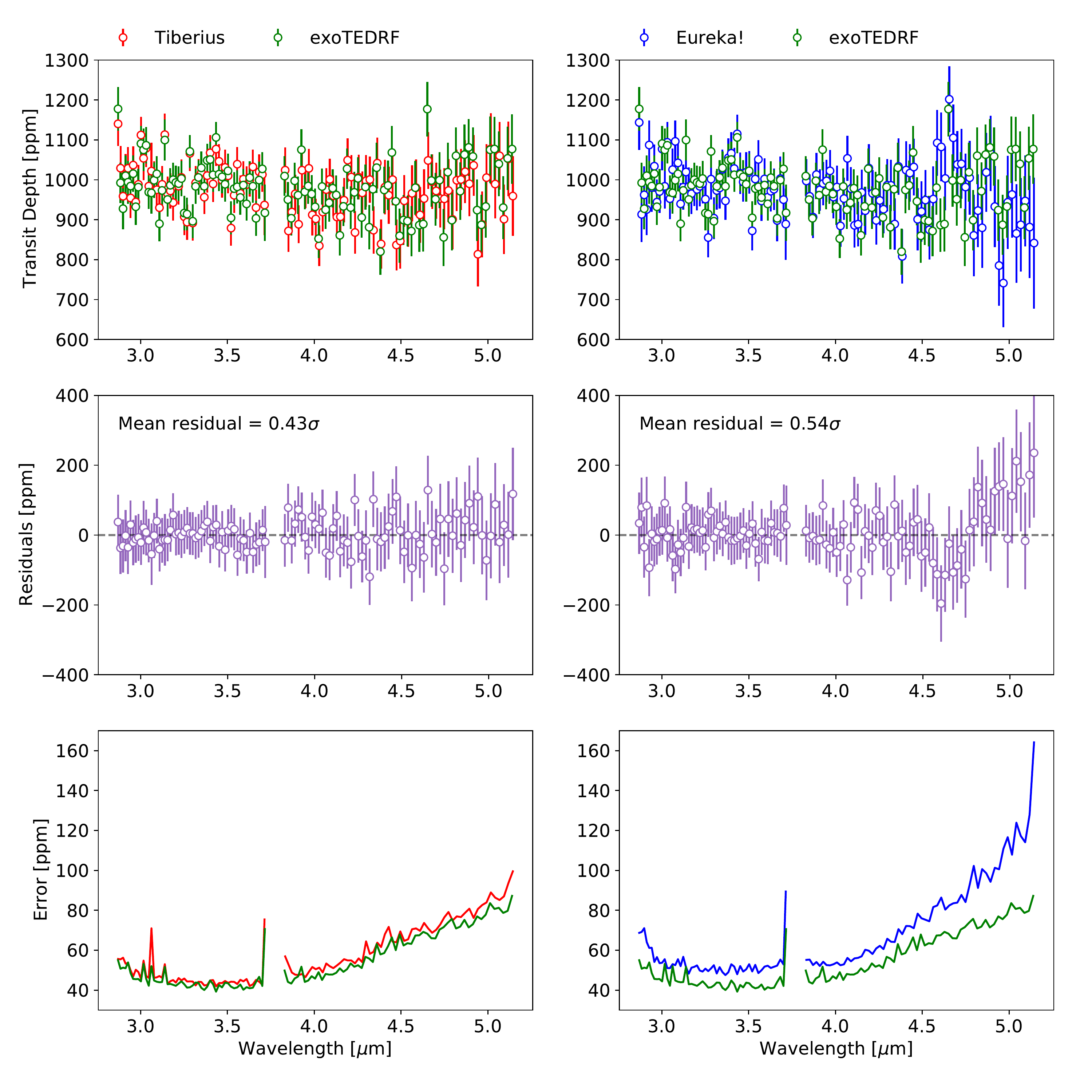}
\end{center}
\vspace{-5mm}\caption{\textbf{Top:} Comparison between the transmission spectra obtained from the \texttt{Tiberius} (red) and \texttt{exoTEDRF} (green) reductions (left), and the \texttt{Eureka!} (blue) and \texttt{exoTEDRF} reductions (right). \textbf{Middle:} Residuals between the \texttt{Tiberius} and \texttt{exoTEDRF} transmission spectra (left), and the \texttt{Eureka!} and \texttt{exoTEDRF} transmission spectra (right). The error bars for the residuals are computed by summing the error bars of the two spectra in quadrature. The mean values of the residuals, normalized by their error bars, are indicated in the panels. The \texttt{Tiberius} and \texttt{exoTEDRF} spectra show great consistency across the full wavelength range, with residuals well centered around 0. As for the \texttt{Eureka!} spectrum, it is generally consistent with \texttt{exoTEDRF} over the NRS1 detector, but there is a significant deviation from 0 at wavelengths past 4.5\,$\mu$m between the two reductions. \textbf{Bottom:} Transit depth errors for the three spectra.} 
\label{fig:reducs_specs}
\end{figure*}

\begin{figure*}[h]
\begin{center}
\includegraphics[width=0.8\textwidth]{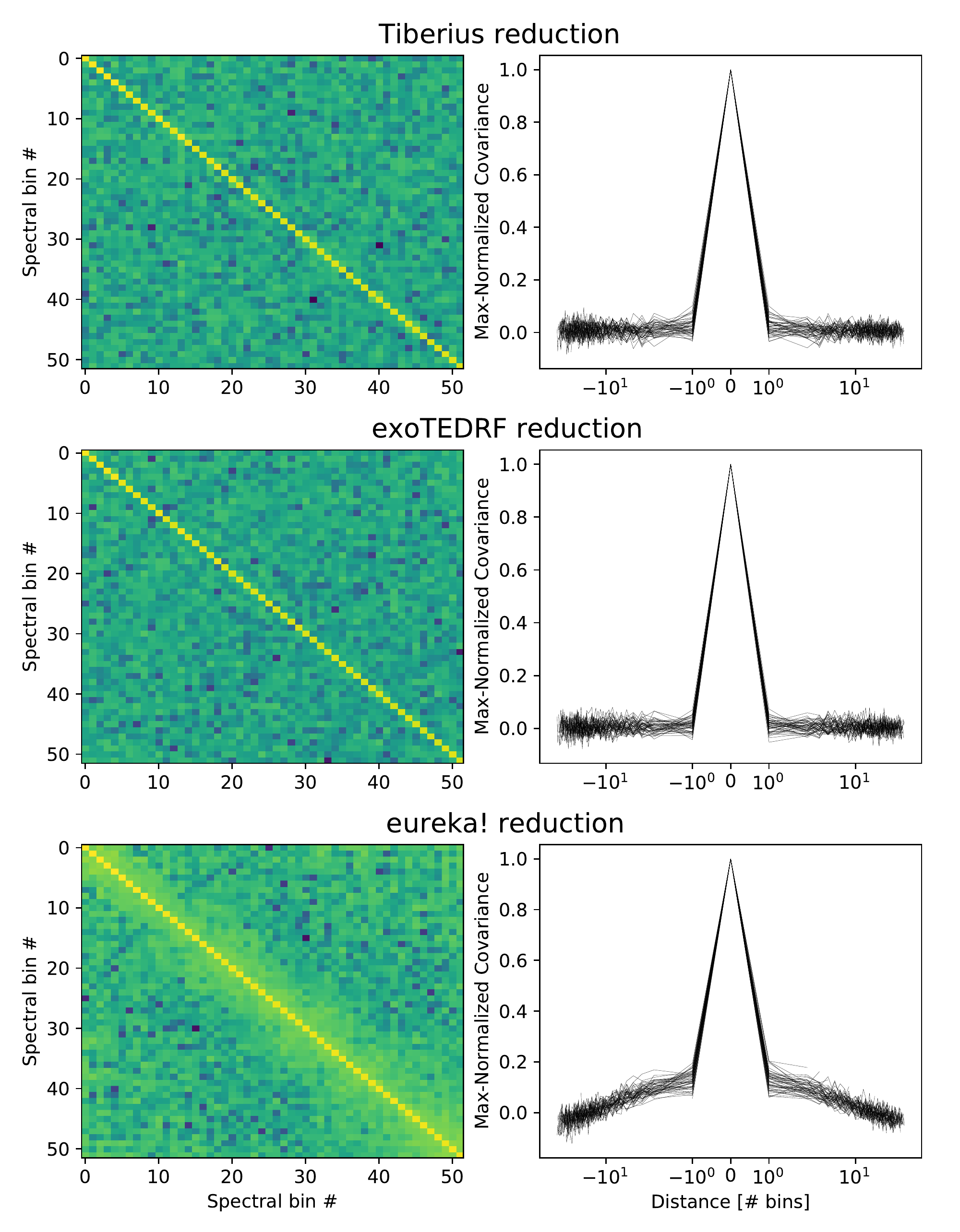}
\end{center}
\vspace{-5mm}\caption{\textbf{Left:} Logarithm of the absolute values of the covariance matrices computed from the residuals of the 51 NRS1 spectroscopic light-curve fits for the \texttt{Tiberius} (top), \texttt{exoTEDRF} (middle), and \texttt{Eureka!} (bottom) reductions. \textbf{Right:} Maximum-normalized covariance curves for all individual spectral bins of the NRS1 detector. The \texttt{Eureka!} reduction shows significant covariance between neighboring spectral bins (away from the diagonal), contrary to the \texttt{Tiberius} and \texttt{exoTEDRF} reductions.} 
\label{fig:spec_corr}
\end{figure*}

\begin{figure}
\begin{center}
\includegraphics[width=0.48\textwidth]{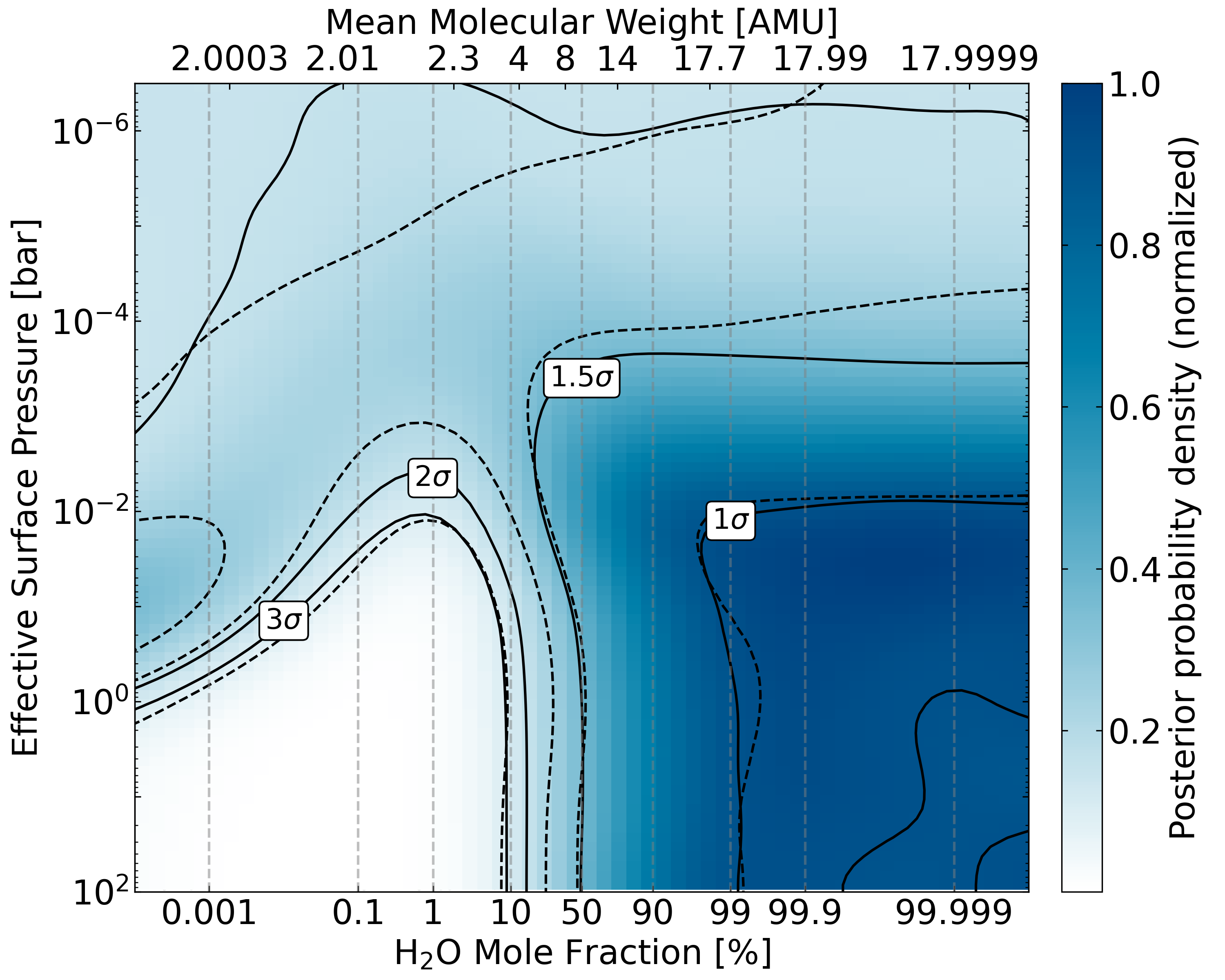}
\end{center}
\vspace{-5mm}\caption{Constraints on the water mole fraction and effective surface pressure from the free chemistry retrieval without TLS and including an offset $\delta_\mathrm{NRS1,2}$. Color represents the normalized posterior probability density from the retrieval assuming no TLS effect, and the full black lines indicate the 1, 1.5, and 2$\sigma$ confidence intervals. The dashed lines represent those same confidence intervals for the retrieval where we assume as a prior the TLS parameter constraints from the retrieval on TOI-270\,d.} 
\label{fig:h2o_vs_pcloud_offset}
\end{figure}

\begin{figure*}
\begin{center}
\includegraphics[width=0.32\textwidth]{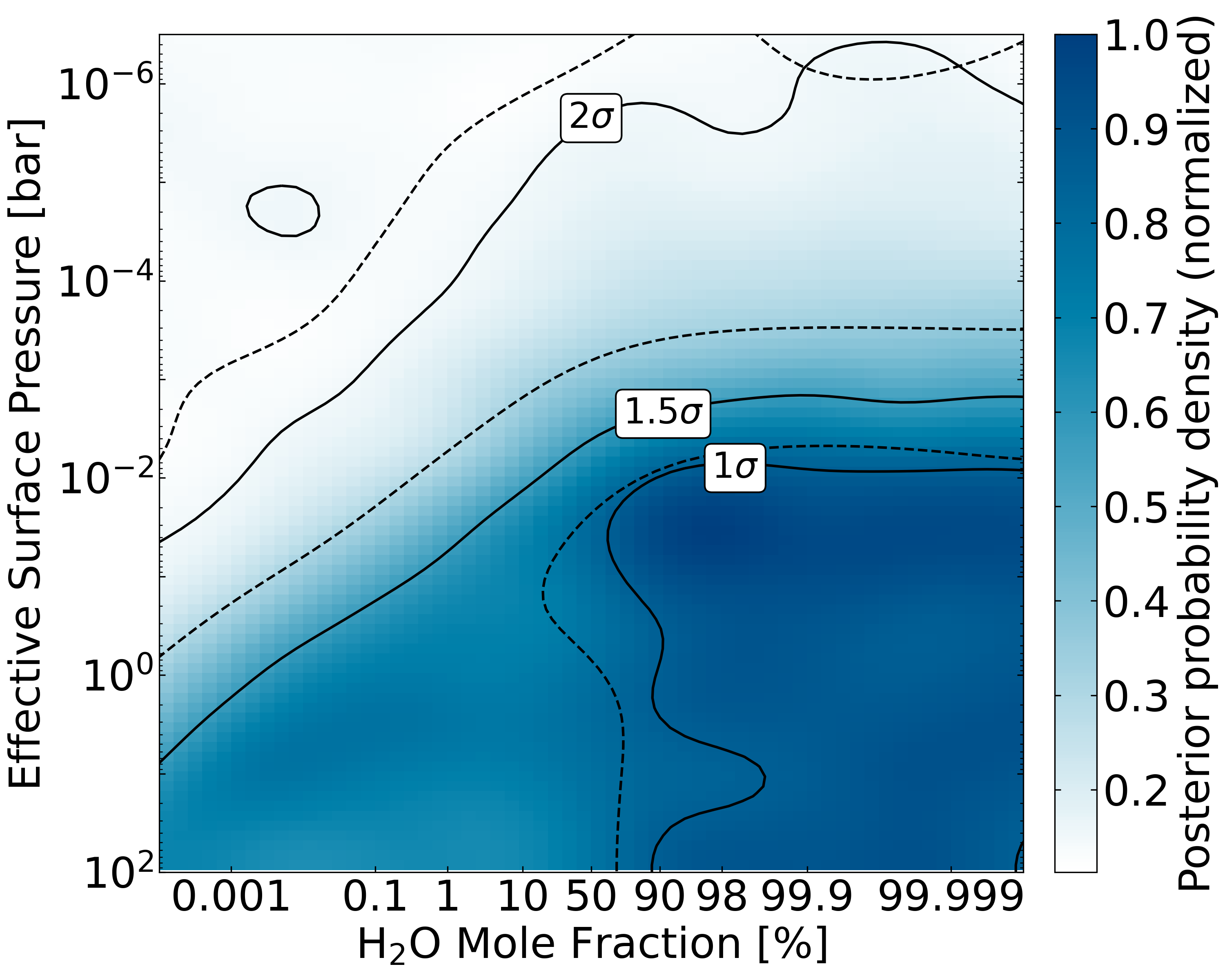}
\includegraphics[width=0.32\textwidth]{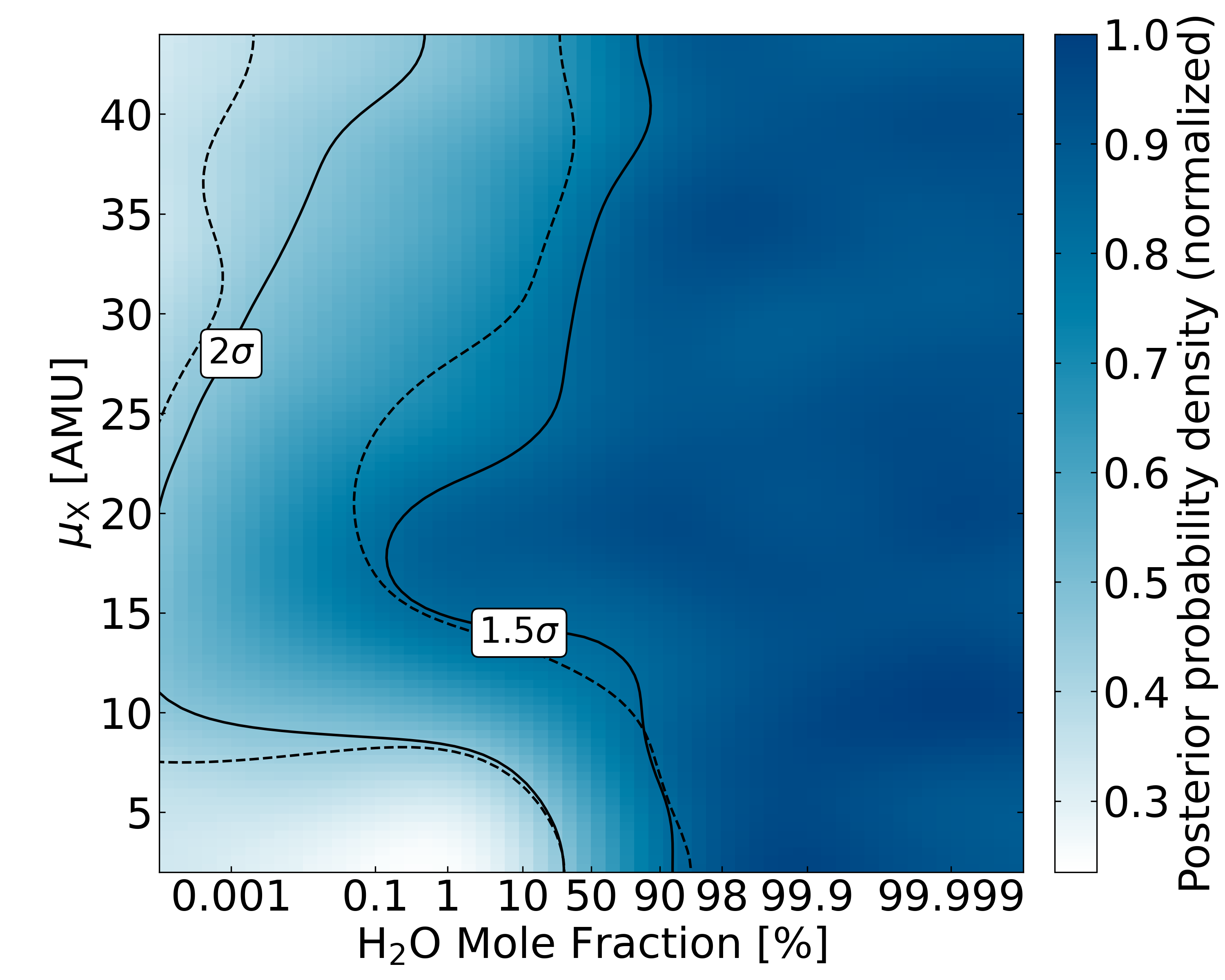}
\includegraphics[width=0.32\textwidth]{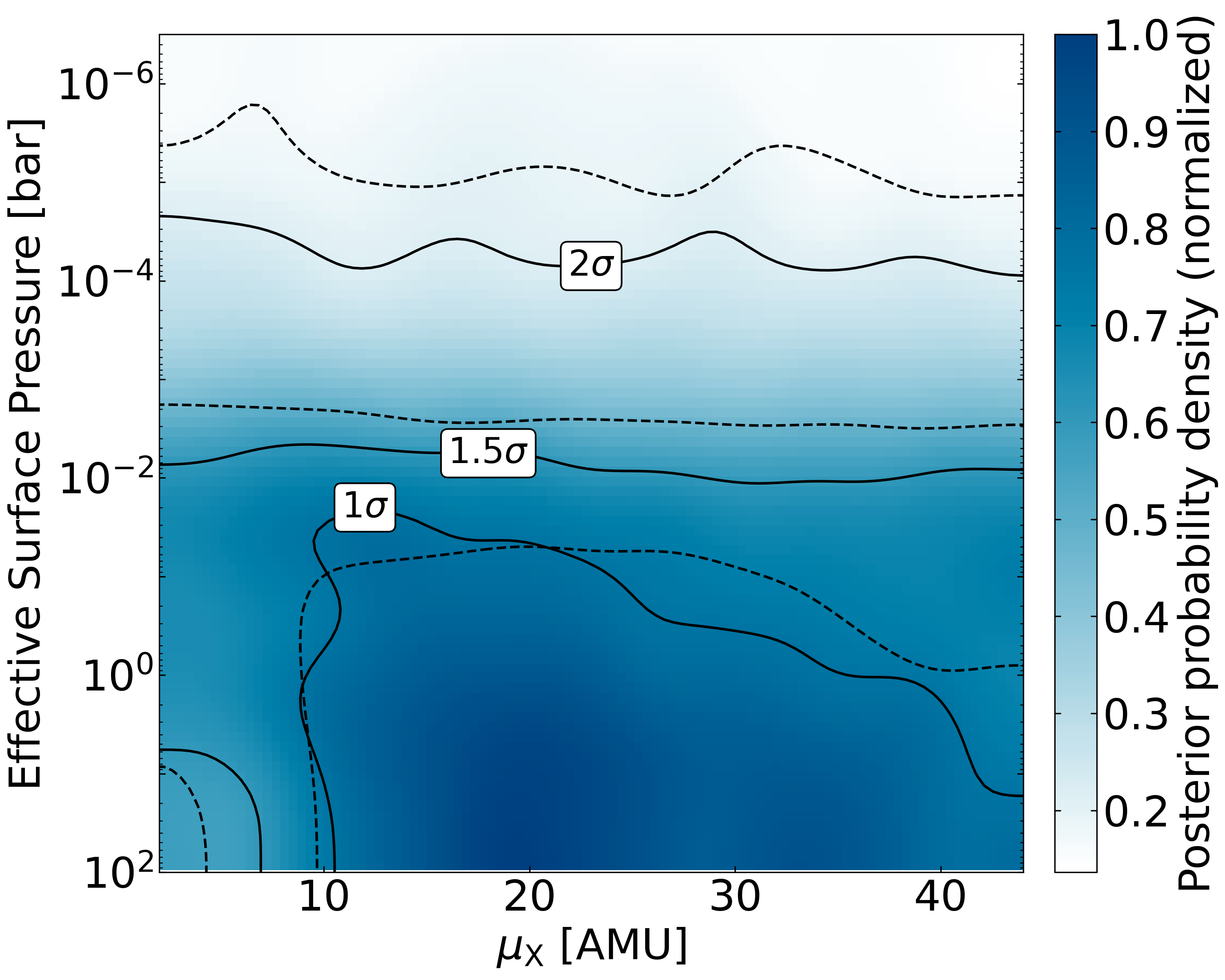}
\end{center}
\vspace{-5mm}\caption{Constraints from the atmospheric retrieval considering the fictitious species X as the background gas and including an offset $\delta_\mathrm{NRS1,2,}$. \textbf{Left:} Joint constraints on the water mole fraction and effective surface pressure, where color represents the normalized probability density. The 1, 1.5, and 2$\sigma$ probability contours are indicated by the full and dashed black lines for the retrievals assuming no TLS and with constrained TLS, respectively. \textbf{Middle:} Joint constraints on the water mole fraction and background gas MMW ($\mu_\mathrm{X}$). The 1.5 and 2$\sigma$ probability contours are indicated by the full and dashed black lines for the retrievals assuming no TLS and with constrained TLS, respectively. Beyond water mole fractions of 90\%, the background MMW is unconstrained as the global MMW is dominated by that of H$_2$O. Below that, the 1.5 and 2$\sigma$ confidence regions are confined to a given range of MMW values, as global MMW that are too high/low would result in water absorption features that are too large/small compared to the tentative feature. \textbf{Right:} Joint constraints on the background gas MMW and effective surface pressure.} 
\label{fig:h2o_vs_pcloud_muX_offset}
\end{figure*}

\begin{table}[h]
\caption{Stellar abundances derived from the NIRPS near-infrared high-resolution spectrum of TOI-270 using PHOENIX ACES models \citep{Husser_2013} with fixed $\log g$ of 5.0\,dex and $T_\mathrm{eff}$ of 3,500\,K, as measured following the methodology of \citet{jahandarComprehensiveHighresolutionChemical2024, jahandarChemicalFingerprintsDwarfs2025}. For each element, the relative-to-solar abundance [X/H] is given by averaging the abundance of all its lines. All elements show abundances that are consistent with solar within 3$\sigma$, with the exception of Si which appears to be an outlier. The overall metallicity [M/H] is defined here as the average abundance of all the measured elements with uncertainty defined as the standard deviation of all values divided by $\sqrt{N-1}$, $N$ being the number of different elements.}
\vspace{3mm}
\centering
\begin{tabular}{lcc}
\hline
\hline
Element & $[$X/H$]$ & No. of lines \\[0.05cm]
\hline
Fe I & $-0.02\pm0.08$ & 18 \\
Mg I & $-0.03\pm0.06$ & 4  \\
Si I & $0.23\pm0.01$ & 3  \\
Ca I & $-0.50\pm0.20$ & 1  \\
Ti I & $-0.01\pm0.06$ & 19 \\
Al I & $-0.02\pm0.23$ & 4  \\
Na I & $-0.35\pm0.20$ & 2  \\
C I  & $-0.13\pm0.56$ & 3  \\
K I  &  $-0.18\pm0.20$ & 4  \\
O I$^*$ & $-0.31\pm0.04$ & 29 \\[0.05cm]
\hline
[M/H] & $-0.13\pm0.07$ & -- \\[0.05cm]
\hline
$^*$Inferred from the OH lines
\label{table:stellarparams}
\end{tabular}
\end{table}

\clearpage

\bibliography{refs}{}
\bibliographystyle{aasjournal}

\end{document}